\documentclass[a4paper, 12pt]{article}

\usepackage[a4paper, margin=1in]{geometry}
\usepackage{amsmath}
\usepackage{amssymb}
\usepackage{graphicx}
\usepackage[onehalfspacing]{setspace}
\usepackage{booktabs}
\usepackage{caption}
\usepackage{appendix}
\usepackage[hidelinks]{hyperref}
\usepackage{natbib}
\setcitestyle{authoryear,open={(},close={)}}
\usepackage{array}
\usepackage{float}
\usepackage{tabularx}
\usepackage{longtable}
\usepackage[flushleft]{threeparttable}

\begin{document}

\bibliographystyle{apalike}

\title{Taming the Option Factor Zoo: A High-Dimensional Analysis}

\author{
    Alexander Walter\textsuperscript{$\dagger$},
    Lukas Zimmer\textsuperscript{$\mathsection$},
    Maxim Ulrich\textsuperscript{$\ddagger$}}
\date{
    \vspace{1em}
    \begin{center}
        \small
        \textsuperscript{$\dagger$} \texttt{alexander.walter@partner.kit.edu} (Corresponding author) \\
        \textsuperscript{$\mathsection$} \texttt{lukas.zimmer@kit.edu}\\
        \textsuperscript{$\ddagger$} \texttt{maxim.ulrich@kit.edu}
    \end{center}
    \vspace{1em}
    \today
}

\maketitle

\begin{abstract}
\noindent Option-implied factors are largely, but not entirely, spanned by the equity factor zoo. We construct 137 option-implied characteristics for optionable U.S.\ stocks from 2004 to 2023, screen them to 20 representative long--short factors, and test each against 160 equity factors with the double-selection LASSO of Feng, Giglio, and Xiu (2020). In-sample, the equity factors account for about 91\% of the option factors' variance, and adding the option factors does not significantly raise the maximum Sharpe ratio. Individually, however, 8 of the 20 option factors, mainly measures of jump tails, kurtosis, and implied-volatility convexity, have non-zero SDF loadings after controlling for the equity zoo, and 4 remain significant after a Bonferroni correction. Options thus add a few pricing-relevant dimensions rather than a new factor zoo.
\end{abstract}
\vspace{1cm}

\newpage

\section{Introduction} \label{sec:intro}

Hundreds of characteristics have been proposed to explain the cross-section of stock returns, and the resulting factor zoo has prompted calls for stricter statistical standards \citep{cochrane2011presidential, harvey2016}. \citet{feng2020taming} provide a tool for one part of this problem: a double-selection LASSO (DS-LASSO) test of whether a new factor contributes to the stochastic discount factor (SDF) once a high-dimensional set of existing factors is controlled for. Almost all factors in the zoo are sorted on characteristics built from past prices and accounting data. Option prices offer a different source of information. They are forward-looking and reveal the risk-neutral distribution of each stock's future return, including its tails. A large literature shows that option-implied measures predict the cross-section of stock returns. Whether factors sorted on these measures add anything once the equity factor zoo is taken into account is the question of this paper.

We build 137 option-implied characteristics for optionable U.S.\ stocks from 2004 to 2023. They cover risk-neutral moments, jump and tail measures, implied-volatility surface shape, and option-implied estimates of the physical return distribution. From each characteristic we form a long--short factor on S\&P 500 constituents, using the same construction as for 152 equity factors based on the characteristics of \citet{jensen2023replication}. Together with eight standard factors from Kenneth French's data library, these form a benchmark of 160 equity factors. Because the option characteristics are highly redundant (91\% of them have a variance inflation factor above 10), we first screen them with gradient-boosted regression trees and SHAP attributions to 20 representative factors. We then test each of the 20 individually against the 160 equity factors with the DS-LASSO. Finally, we ask whether the option factors as a group expand the investment opportunity set and how much of their variation the equity factors share.

Our answer has two parts. First, the option factor zoo is largely redundant. In-sample, the 160 equity factors account for 91\% of the variance of the 20 option factors, compared with about 69\% for pure noise of the same dimensions. The option factors, in turn, account for 35\% of the equity factors' variance, compared with about 9\% for noise. A GRS test does not reject that the option factors' alphas on the equity factors are jointly zero ($p = 0.25$). Adding the option factors raises the in-sample maximum Sharpe ratio from 3.35 to 3.74, but this increase is not statistically significant. Second, the redundancy is not complete. Eight of the 20 option factors have SDF loadings that are significantly different from zero at the 5\% level after controlling for the equity zoo. Four of them remain significant after a Bonferroni correction for testing 20 factors: implied-volatility convexity, the share of risk-neutral variance due to large jumps, left-tail jump variation, and risk-neutral kurtosis. Measures of tail size, kurtosis, and convexity carry positive loadings; right-tail jump intensity and option-implied physical variance carry negative loadings.

These results are in-sample and conditional on the screening step, and we discuss both caveats in detail. The specific factors selected by the screen vary across option maturities and subsamples, because the screen picks one representative from groups of near-duplicate measures. The aggregate conclusions do not vary: across seven option maturities, the equity factors explain 91\% to 94\% of the option factors' variance, and the GRS test rejects only at the longest maturity.

We contribute to three strands of literature. First, we add to the work on the factor zoo \citep{harvey2016, feng2020taming, jensen2023replication} by applying a high-dimensional test of SDF contributions to an entire family of option-implied factors. Second, we add to the literature on option-implied information and the cross-section of stock returns \citep[e.g.,][]{bali2009volatility, conrad2013ex, an2014joint, bali2024search, neuhierl2023option}. \citet{neuhierl2023option} show that option characteristics predict stock returns beyond firm characteristics. We ask a complementary question: whether factors sorted on option characteristics contribute to the SDF once the equity factor zoo is controlled for. Our evidence suggests that most of their information is already reflected in equity factors, and that the incremental part is concentrated in tail and higher-moment measures. Third, we provide a dataset of 137 option-implied characteristics, including option-implied estimates of physical moments and their systematic and idiosyncratic components.

The remainder of the paper is organized as follows. Section~\ref{sec:lit} reviews related literature. Section~\ref{sec:data} describes the data and the construction of factors and test assets. Section~\ref{sec:method} describes the screening step, the DS-LASSO test, and the aggregate tests. Section~\ref{sec:findings} presents the results, Section~\ref{sec:discussion} discusses them together with the limitations of our design, and Section~\ref{sec:conclusion} concludes.

\subsection{Related Literature} \label{sec:lit}

\noindent\textbf{The factor zoo.} The CAPM was challenged by the three-factor model of \citet{fama1993common}, extended with momentum by \citet{carhart1997persistence}. Later models added profitability and investment \citep{hou2015digesting, fama2015five}, and factors motivated by funding constraints \citep{frazzini2014betting} or mispricing \citep{stambaugh2017mispricing}. The number of proposed factors grew into what \citet{cochrane2011presidential} called a factor zoo. \citet{harvey2016} argue that the multiple testing implicit in this search requires a $t$-statistic hurdle of about 3.0 for new factors, and \citet{chordia2020anomalies} reach a similar conclusion. \citet{hou2020replicating} and \citet{jensen2023replication} study how many anomalies replicate; we build our equity factors from the characteristics of \citet{jensen2023replication}.\\

\noindent\textbf{High-dimensional methods.} When the number of candidate factors is large relative to the time series, standard two-pass regressions suffer from omitted-variable bias and model-selection mistakes. One group of methods builds a parsimonious SDF or predicts returns: \citet{kozak2020shrinking} shrink the SDF towards high-variance principal components, \citet{gu2020empirical} compare machine-learning methods for return prediction, \citet{freyberger2020dissecting} select characteristics nonparametrically, and \citet{lettau2020factors} and \citet{lettau2023high} use risk-premium PCA and tensor factor models, respectively. Another group focuses on inference. \citet{bryzgalova2023bayesian} average over a very large number of models with a Bayesian approach that is robust to weak factors \citep{bryzgalova2015spurious}. \citet{giglio2021asset} estimate risk premia when some factors are omitted, using a three-pass method. Our test follows \citet{feng2020taming}, who adapt the double-selection LASSO of \citet{belloni2014inference} to ask whether a new factor's SDF loading is non-zero after controlling for a high-dimensional set of existing factors.\\

\noindent\textbf{Option-implied information.} \citet{bakshi2003stock} show how to extract model-free risk-neutral moments from option prices. \citet{bollerslev2009expected} find that the variance risk premium predicts market returns, and at the firm level \citet{bali2009volatility} relate the spread between implied and realized volatility to future stock returns. \citet{cremers2010deviations}, \citet{xing2010what}, and \citet{an2014joint} document that deviations from put--call parity, the implied-volatility smirk, and changes in implied volatility predict stock returns. For risk-neutral skewness, the evidence is mixed. \citet{conrad2013ex} find that stocks with more negative risk-neutral skewness earn higher subsequent returns, whereas \citet{rehman2012risk} and \citet{stilger2017what} find a positive relation and attribute the difference to how skewness is measured. \citet{bollerslev2015efficient} decompose option-implied variance into jump and tail components and show that tail risk premia predict returns. \citet{neuhierl2023option} study a large set of option characteristics as cross-sectional predictors. Closest to our aggregate question, \citet{bali2024search} propose a factor model for the cross-section of optionable stocks, with factors based on the spread between implied and realized volatility, the call--put implied-volatility spread, and the difference between changes in call and put implied volatilities. A separate literature builds factor models for option returns rather than stock returns \citep{buchner2022factor, horenstein2026common}. We study stock returns throughout.

\section{Data} \label{sec:data}

\subsection{Sample and Universes} \label{sec:sample}

Our sample runs from January 2004 to July 2023. The first month is used to construct lagged variables, so returns are evaluated from February 2004 to July 2023, a total of 234 months.\footnote{We also attempted the analysis at a daily frequency. The noise in daily returns led to near-singular covariance matrices, and the DS-LASSO estimation did not converge.} Equity characteristics come from the dataset of \citet{jensen2023replication}, which builds on the anomaly list of \citet{hou2020replicating} and uses CRSP and Compustat data. Option-implied characteristics are constructed from CBOE option data. Because option-implied characteristics exist only for stocks with traded options, we restrict the analysis to stocks for which both the equity characteristics and sufficient option data are available.

Three universes are used. Both the option factors and the equity factors are formed from the S\&P 500 constituents in this intersected universe, with index membership updated monthly. The test assets are formed from all stocks in the intersected universe. Finally, we obtain the Fama--French five factors \citep{fama2015five}, the momentum factor \citep{carhart1997persistence}, the short- and long-term reversal factors, and the corresponding benchmark portfolios from Kenneth French's data library. These are built from the full CRSP universe, a difference we accept in exchange for using the standard benchmark series. Table~\ref{tab:data_overview} summarizes the sample.

\begin{table}[H]
\centering
\caption{\textbf{Sample Selection}}
\label{tab:data_overview}
\begin{tabular}{lc}
\toprule
 & \textbf{Value} \\
\midrule
\multicolumn{2}{l}{\textit{Panel A: Time dimension}} \\
Return sample & Feb 2004 -- Jul 2023 \\
Number of months & 234 \\
Frequency & Monthly \\
\midrule
\multicolumn{2}{l}{\textit{Panel B: Cross-section}} \\
Stocks in the equity universe & 17,586 \\
Stocks with listed options & 9,987 \\
Final universe (sufficient option data) & 7,126 \\
\midrule
\multicolumn{2}{l}{\textit{Panel C: Observations and portfolio size}} \\
Firm-month observations & 824,081 \\
Avg.\ stocks per month in each leg of a factor portfolio & 42 \\
Avg.\ stocks per month in each leg of a test-asset portfolio & 369 \\
\bottomrule
\end{tabular}
\begin{flushleft}
\footnotesize This table describes the sample. Raw data start in January 2004; the first month is used to construct lagged variables. Panel B shows the reduction from the equity universe to stocks with listed options and to the final universe with sufficient option data to estimate the risk-neutral density. The panel is unbalanced. Panel C reports the number of firm-month observations and the average number of stocks per month in the high and low legs of the factor and test-asset portfolios.
\end{flushleft}
\end{table}

\subsection{Option-Implied Characteristics} \label{sec:optchar}

We first estimate each stock's risk-neutral density (RND) from option prices, following \citet{breeden1978prices}. To obtain a smooth and differentiable implied-volatility surface, we use the one-dimensional kernel estimator of \citet{ulrich2023implied}, which smooths along moneyness $m_i = K_i/F_i$ and reduces cross-validation errors relative to standard approaches. Implied volatilities are extrapolated linearly beyond the observed strikes, which stabilizes the tails of the density. From these densities and the underlying option prices we compute characteristics from the following groups:

\begin{itemize}
    \item \textbf{Implied moments:} risk-neutral volatility, skewness, and kurtosis, following the model-free approach of \citet{bakshi2003stock}.
    \item \textbf{Jump and tail measures:} decompositions of risk-neutral quadratic variation (QV) into continuous and jump components, left- and right-tail jump variation, and jump intensities, following \citet{bollerslev2015efficient}.
    \item \textbf{Disaster risk:} the rare-disaster index of \citet{gao2018disaster}.
    \item \textbf{Implied-volatility surface shape:} the smirk, convexity, and slopes of the out-of-the-money wings.
    \item \textbf{Other measures:} option-implied betas, option-implied equity premia, and the SVIX of \citet{martin2017expected}.
\end{itemize}

In addition, we construct option-implied measures of the physical ($\mathbb{P}$) return distribution, labeled P\_implQ. For each stock, we compute the ratio of realized high-frequency physical moments to the corresponding one-day risk-neutral moments, $\mathbb{P}/\mathbb{Q}_{1d}$. We then scale the risk-neutral moments at longer maturities $\tau$ by this ratio, $\mathbb{P}_{\tau} = (\mathbb{P}/\mathbb{Q}_{1d}) \times \mathbb{Q}_{\tau}$. By construction, the one-day physical distribution matches the high-frequency sample paths. Carrying a constant $\mathbb{P}/\mathbb{Q}$ ratio to longer maturities is a strong assumption, and we treat these measures as approximations of physical moments. We further decompose the physical and risk-neutral moments into systematic (suffix \_sys) and idiosyncratic (suffix \_eps) components. The variance decomposition is additive; the decompositions of the higher, normalized moments are not.

Appendix~B lists all option-implied characteristics with their sources. It contains 138 entries, one of which, the jump cutoff $k_t$ of \citet{bollerslev2015efficient}, is an input to other measures rather than a signal; the analysis uses the remaining 137.

\subsection{Factors and Test Assets} \label{sec:factors}

All option-implied characteristics are computed at a constant maturity of 30 days ($\tau = 30$), which matches the monthly rebalancing frequency; Section~\ref{sec:robustness} reports results for maturities from 1 to 60 days. At the end of each month, we sort the S\&P 500 constituents in our universe into deciles on a characteristic and form a factor that is long the top decile and short the bottom decile. Portfolios are value-weighted and held for the following month. Our equity benchmark consists of 160 factors. Of these, 152 are formed in the same way from the characteristics of \citet{jensen2023replication} listed in Appendix~A.\footnote{Appendix~A lists 153 characteristics. We exclude price per share (prc) because it has no clear economic interpretation as a sorting variable.} The remaining eight are the factors from Kenneth French's data library listed in Section~\ref{sec:sample}: the five Fama--French factors, momentum, and short- and long-term reversal. We report gross returns and ignore transaction costs throughout. Excess returns are computed with the three-month Treasury-bill rate from FRED, converted to a monthly rate and lagged by one month. As a robustness check, we use the option-implied risk-free rate of \citet{vanbinsbergen2022risk}, as published by \citet{vanbinsbergen2023options} (TOPT).

The DS-LASSO test requires test assets. Following \citet{feng2020taming}, we form, for each characteristic, six portfolios from independent $3 \times 2$ sorts on the characteristic and market capitalization, using all stocks in the intersected universe rather than only S\&P 500 constituents. We add the benchmark portfolios from Kenneth French's data library that correspond to the factors listed above.

\section{Methodology} \label{sec:method}

Our analysis has three steps. We first reduce the 137 option factors to a set of representatives (Section~\ref{sec:screen}). We then test each representative individually against the equity factors with the DS-LASSO (Section~\ref{sec:dslasso}). Finally, we evaluate the option factors as a group, asking whether they expand the investment opportunity set and how much variation they share with the equity factors (Section~\ref{sec:big_picture_method}). All three steps are in-sample.

\subsection{Screening the Option Factors} \label{sec:screen}

The option factors are highly collinear: 91\% of them have a variance inflation factor (VIF) above 10 and 62\% above 100. Many are close variants of the same measure, for example the same jump variation computed with cutoffs of 7.5\%, 10\%, and 20\%. Because the DS-LASSO tests one option factor at a time against the equity controls, this collinearity does not affect any single test. It does make testing all 137 factors unattractive. Near-duplicate factors produce near-identical tests, which add little information but inflate the number of tests and therefore the multiple-testing burden.\footnote{An initial attempt to include all 137 option factors together with the equity factors in a single estimation led to near-singular covariance matrices.} We therefore select a smaller set of representatives before testing.

We use gradient-boosted regression trees (GBR) with SHAP attributions for this screen. The target is the equal-weighted average return of all test-asset portfolios in month $t$, and the features are the returns of the 137 option factors in the same month. The screen is thus a contemporaneous fit rather than a forecast: it asks which option factors move most with the average test asset. Unlike principal components, which summarize the factors without reference to returns, this criterion selects actual factors and ranks them by their relevance for the test assets. It also favours factors with large exposure to the market, which we take into account when interpreting the results.

We use the histogram-based gradient-boosting regressor with a learning rate of 0.01, at most 100 boosting iterations, and a maximum tree depth of 3, which limits overfitting given the short time series. We evaluate the fit with a five-fold walk-forward time-series split (expanding windows) and, in addition, five rolling windows. For the final selection, the model is estimated on the full sample. SHAP values \citep{lundberg2017unified} decompose each fitted value into contributions of the individual factors, and we rank factors by their mean absolute SHAP value. We keep the largest set of top-ranked factors whose cumulative share of total attribution does not exceed 90\%. This yields 20 factors, which account for 89.3\% of the total attribution.

Using the full sample for the screen and then testing the selected factors on the same sample means that our test results are conditional on the screen. The screen is based on a criterion (co-movement with the average test asset) that differs from the one tested later (a non-zero SDF loading given the equity controls), but it is not independent of it. Section~\ref{sec:screen_results} reports how stable the selection is across maturities and subsamples.

For the 20 selected factors, we report annualized means, volatilities, and Sharpe ratios, and we estimate alphas relative to the Fama--French five-factor model augmented with momentum. We also run the GRS test \citep{gibbons1989test} of whether the alphas of the 20 option factors relative to the 160 equity factors are jointly zero.

\subsection{Double-Selection LASSO} \label{sec:dslasso}

We test whether an option factor $g_t$ contributes to the SDF after controlling for the equity factors $h_t$, following \citet{feng2020taming} and using their code without modification. We test each of the 20 option factors individually. Denote the vector of average returns of the $n$ test assets by $\bar r$, and the $n \times 1$ and $n \times p$ matrices of sample covariances of the test assets with $g_t$ and $h_t$ by $\hat C_g$ and $\hat C_h$. The model is
\begin{equation}
\bar r = \gamma_0 + \hat C_g \lambda_g + \hat C_h \lambda_h + \varepsilon,
\end{equation}
where $\lambda_g$ and $\lambda_h$ are SDF loadings. The null hypothesis is $\lambda_g = 0$: once the equity factors are accounted for, the option factor is not needed to price the test assets. Because most of the $p$ controls are irrelevant for any given factor, and a naive selection of controls causes omitted-variable bias, the procedure selects the controls in two steps.

\noindent\textbf{Step 1 (pricing relevance).} A cross-sectional LASSO regression of $\bar r$ on $\hat C_h$ selects the equity factors that help price the test assets, collected in the set $I_1$.

\noindent\textbf{Step 2 (confounding).} A cross-sectional LASSO regression of $\hat C_g$ on $\hat C_h$ selects the equity factors whose covariances with the test assets are correlated with those of the option factor, collected in $I_2$. Omitting these factors would bias the estimate of $\lambda_g$.

\noindent\textbf{Estimation.} $\lambda_g$ is estimated by an OLS regression of $\bar r$ on $\hat C_g$ and $\hat C_{h,I}$, where $I = I_1 \cup I_2$.

\noindent\textbf{Inference.} For the standard errors, the implementation of \citet{feng2020taming} runs a third, time-series LASSO of $g_t$ on $h_t$, with the tuning parameter chosen by $K$-fold time-series cross-validation and the one-standard-error rule. The residual
\begin{equation}
z_t = g_t - \Pi_{I_3} g_t,
\end{equation}
where $\Pi_{I_3}$ denotes the projection on the selected controls, enters the heteroskedasticity-robust estimate of the asymptotic variance of $\hat\lambda_g$. This accounts for the model-selection steps in the inference.

The tuning parameters of steps 1 and 2 are chosen by 10-fold cross-validation, averaged over multiple random seeds. Following \citet{feng2020taming}, we report $\lambda_s$, the SDF loading scaled to correspond to a unit univariate beta exposure. It can be read as the average excess return, in basis points per month, of a portfolio with a unit univariate beta with respect to the factor, holding exposures to the selected controls fixed. A positive $\lambda_s$ means that high realizations of the factor coincide with states of low marginal utility.

$\lambda_s$ is not the same as the factor's average return. For a traded factor, the average return is its risk premium, which reflects its exposure to all priced sources of risk, including the equity factors. $\lambda_s$ measures only the factor's marginal contribution to the SDF, given the controls. The two can differ in sign when a factor is correlated with priced equity factors. Following \citet{feng2020taming}, $\lambda_s$ is estimated from the cross-section of test portfolios sorted on the characteristics, including the option-implied ones, rather than from the time series of the long--short factor itself. We report both quantities side by side, as \citet{feng2020taming} do.

Because we run one test per factor, 20 tests in total, some rejections at the 5\% level are expected by chance even if no option factor matters: the probability of at least one false rejection is $1 - 0.95^{20} \approx 64\%$. We therefore also report whether each result survives a Bonferroni correction (a two-sided 5\% test at level $0.05/20$, i.e., $|t| > 3.02$) and the less conservative procedure of \citet{benjamini1995controlling}, which controls the expected share of false discoveries at 5\%. The Bonferroni threshold is close to the hurdle of $t > 3.0$ proposed by \citet{harvey2016}.

\subsection{Aggregate Tests} \label{sec:big_picture_method}

\noindent\textbf{Investment opportunity set.} We compute the maximum Sharpe ratio attainable from the equity factors, from the option factors, and from both combined. Tangency-portfolio weights are proportional to the inverse covariance matrix times the vector of mean excess returns. Because the combined set contains 180 factors and the sample only 234 months, we use the shrinkage covariance estimator of \citet{ledoit2004well}. We report annualized Sharpe ratios. These in-sample maximum Sharpe ratios are biased upward, and the bias grows with the number of assets \citep{kan2024insample}, so adding factors raises the in-sample value even when they carry no information. The formal test of whether the option factors improve the maximum Sharpe ratio is the GRS test described above: its statistic is a function of the difference between the squared maximum Sharpe ratios with and without the option factors \citep{gibbons1989test, barillas2017which}.

\noindent\textbf{Shared variation.} We measure how much variation the two sets of factors share with canonical correlation analysis (CCA) and the redundancy index of \citet{stewart1968general}. For a target set $Y$ and a source set $X$,
\begin{equation}
RI_{Y|X} = \sum_{i=1}^{k} \overline{L_{Y,i}^2} \times \rho_i^2 ,
\end{equation}
where $\rho_i$ is the $i$-th canonical correlation and $\overline{L_{Y,i}^2}$ is the average squared correlation between the variables in $Y$ and their $i$-th canonical variate. We estimate the CCA on standardized factor returns and use all $k = 20$ canonical pairs. The index has a simple interpretation: with all canonical pairs, $RI_{Y|X}$ equals the average $R^2$ from regressing each variable in $Y$ on all variables in $X$. The statement ``$X$ explains $k$\% of $Y$'' therefore means that, on average, a linear combination of the $X$ factors reproduces $k$\% of each $Y$ factor's monthly variance in-sample. We call $1 - RI_{Y|X}$ the unique share of $Y$. It includes estimation noise and is therefore not automatically economically distinct information. Because in-sample $R^2$ rises mechanically with the number of regressors, we compare each index with its value for independent noise: about $160/233 \approx 0.69$ when the 160 equity factors explain an option factor, and about $20/233 \approx 0.09$ when the 20 option factors explain an equity factor.

\section{Results} \label{sec:findings}

\subsection{Redundancy in the Option Factor Zoo and the Screen} \label{sec:screen_results}

\begin{figure}[H]
    \centering
    \includegraphics[width=0.85\textwidth]{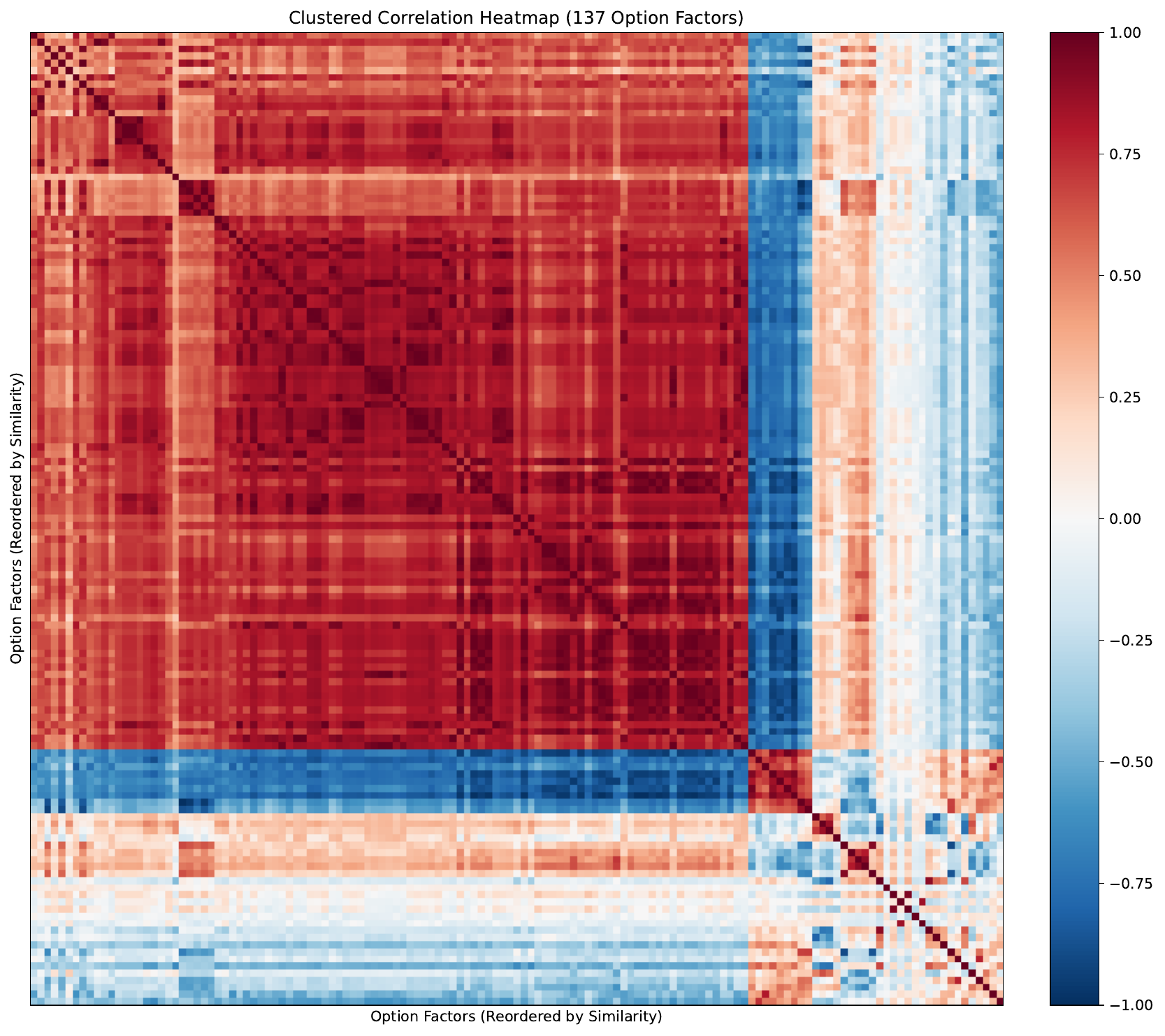}
    \caption{\textbf{Clustered Correlation Matrix of the Option Factors.} The figure shows the correlation matrix of the 137 option factors, reordered by hierarchical agglomerative clustering with Ward's linkage. Dark red blocks along the diagonal are clusters of highly correlated factors.}
    \label{fig:heatmap_plot}
\end{figure}

Figure~\ref{fig:heatmap_plot} shows the correlation matrix of the 137 option factors, ordered by hierarchical clustering. Most factors fall into a few large blocks of highly correlated measures, in line with the VIF statistics reported above.

Table~\ref{tab:gbr_selected_factors} lists the 20 factors retained by the screen, grouped by economic category. Eleven are jump and tail measures of \citet{bollerslev2015efficient} (BTX). The two largest attributions belong to option-implied physical variance (P\_implQ\_var, 14.7\%) and right-tail jump variation (BTX\_RJV\_k, 13.3\%). The selection still contains near-duplicates, such as BTX\_RJV\_k and BTX\_RJV\_10, or BTX\_EQV\_relRightTail\_10 and BTX\_EQV\_relRightTail\_k. The screen therefore prioritizes factors for testing rather than removing all redundancy.

\begin{table}[H]
\centering
\begin{threeparttable}
\caption{\textbf{Option Factors Selected by the Screen}}
\label{tab:gbr_selected_factors}
\begin{small}
\begin{tabularx}{\textwidth}{Xcc}
\toprule
\textbf{Factor} & \textbf{Mean $|$SHAP$|$} & \textbf{Share (\%)} \\
\midrule
\multicolumn{3}{l}{\textbf{Volatility Level \& Dynamics}} \\
\hspace{0.3cm} P\_implQ\_var & 0.005653 & 14.65 \\
\hspace{0.3cm} SVIX2\_put\_annual & 0.002096 & 5.43 \\
\addlinespace
\multicolumn{3}{l}{\textbf{Jump \& Discontinuous Risk}} \\
\hspace{0.3cm} BTX\_RJV\_k & 0.005117 & 13.26 \\
\hspace{0.3cm} BTX\_EQV\_relTail\_k & 0.004854 & 12.58 \\
\hspace{0.3cm} BTX\_LJV\_7\_5 & 0.001249 & 3.24 \\
\hspace{0.3cm} BTX\_MRJI\_20\_annual & 0.000980 & 2.54 \\
\hspace{0.3cm} BTX\_EQV\_relRightTail\_10 & 0.000822 & 2.13 \\
\hspace{0.3cm} BTX\_MLJI\_20\_annual & 0.000742 & 1.92 \\
\hspace{0.3cm} BTX\_EQV\_relRightTail\_k & 0.000630 & 1.63 \\
\hspace{0.3cm} BTX\_MRJI\_k\_annual & 0.000590 & 1.53 \\
\hspace{0.3cm} BTX\_EQCV\_k & 0.000582 & 1.51 \\
\hspace{0.3cm} BTX\_RJV\_10 & 0.000552 & 1.43 \\
\hspace{0.3cm} BTX\_alpha\_plus & 0.000494 & 1.28 \\
\addlinespace
\multicolumn{3}{l}{\textbf{Skewness, Kurtosis \& Asymmetry}} \\
\hspace{0.3cm} bakshi\_X & 0.002656 & 6.88 \\
\hspace{0.3cm} bakshiKurt & 0.000731 & 1.89 \\
\hspace{0.3cm} ivs\_smirk & 0.000620 & 1.61 \\
\hspace{0.3cm} ivs\_convexity & 0.000586 & 1.52 \\
\addlinespace
\multicolumn{3}{l}{\textbf{Systematic vs. Idiosyncratic}} \\
\hspace{0.3cm} P\_implQ\_var\_eps & 0.002368 & 6.14 \\
\hspace{0.3cm} RND\_kurt\_eps & 0.000640 & 1.66 \\
\addlinespace
\multicolumn{3}{l}{\textbf{Other Risk Measures}} \\
\hspace{0.3cm} RND\_mu & 0.002491 & 6.45 \\
\bottomrule
\end{tabularx}
\end{small}
\begin{flushleft}
\footnotesize This table lists the 20 option factors retained by the GBR screen at the 30-day maturity, grouped by economic category. Mean $|$SHAP$|$ is the mean absolute SHAP attribution of the factor in the full-sample GBR fit of the equal-weighted average test-asset return. Share is the factor's percentage of the total attribution across all 137 option factors; the 20 factors together account for 89.3\%. Appendix~B defines all factors.
\end{flushleft}
\end{threeparttable}
\end{table}

Table~\ref{tab:gbr_stability} summarizes how stable the selection is. Five factors are selected at six or seven of the seven maturities, and the selection counts in the expanding windows are higher than in the rolling windows. Several factors that turn out to be significant below, however, are selected only at the 30-day maturity and rarely in subsamples, among them bakshiKurt, RND\_kurt\_eps, and BTX\_MRJI\_20\_annual. We read this as a property of the redundant option zoo: when several measures carry nearly the same information, the screen picks one of them, and which one it picks varies. The selected factors should therefore be read as representatives of groups of related measures (Figure~\ref{fig:heatmap_plot}) rather than as uniquely identified signals. Appendix Tables~\ref{tab:gbr_persistence} and \ref{tab:gbr_window_selections} report the full selection results, and Table~\ref{tab:gbr_rmse} reports the out-of-sample fit of the GBR. Because the target is contemporaneous, the latter measures fit rather than forecasting ability. The root mean squared errors range from 3.5\% to 8.2\% across windows and are nearly identical across maturities.

\begin{table}[H]
\centering
\caption{\textbf{Stability of the Screen}}
\label{tab:gbr_stability}
\begin{small}
\begin{tabular}{lcccc}
\toprule
 & \textbf{Maturities} & \multicolumn{2}{c}{\textbf{Subsamples, 30 days}} & \textbf{All subsamples} \\
\cmidrule(lr){3-4}
\textbf{Factor} & \textbf{(of 7)} & \textbf{Expanding (of 5)} & \textbf{Rolling (of 5)} & \textbf{(of 70)} \\
\midrule
bakshi\_X & 5 & 1 & 1 & 11 \\
bakshiKurt & 1 & 3 & 0 & 3 \\
BTX\_alpha\_plus & 2 & 1 & 0 & 2 \\
BTX\_EQCV\_k & 2 & 2 & 0 & 5 \\
BTX\_EQV\_relRightTail\_10 & 5 & 2 & 0 & 15 \\
BTX\_EQV\_relRightTail\_k & 6 & 1 & 0 & 13 \\
BTX\_EQV\_relTail\_k & 3 & 2 & 2 & 8 \\
BTX\_LJV\_7\_5 & 5 & 2 & 0 & 13 \\
BTX\_MLJI\_20\_annual & 6 & 2 & 0 & 15 \\
BTX\_MRJI\_20\_annual & 1 & 1 & 0 & 1 \\
BTX\_MRJI\_k\_annual & 7 & 2 & 0 & 18 \\
BTX\_RJV\_10 & 1 & 1 & 0 & 1 \\
BTX\_RJV\_k & 7 & 3 & 1 & 28 \\
ivs\_convexity & 6 & 1 & 0 & 7 \\
ivs\_smirk & 1 & 2 & 1 & 6 \\
P\_implQ\_var & 3 & 4 & 2 & 32 \\
P\_implQ\_var\_eps & 2 & 2 & 0 & 18 \\
RND\_kurt\_eps & 1 & 1 & 0 & 3 \\
RND\_mu & 3 & 1 & 0 & 8 \\
SVIX2\_put\_annual & 2 & 1 & 1 & 6 \\
\bottomrule
\end{tabular}
\end{small}
\begin{flushleft}
\footnotesize This table reports how often each of the 20 factors selected at the 30-day maturity is selected in other specifications. Maturities is the number of the seven maturities (1 to 60 days) at which the full-sample screen selects the factor (Table~\ref{tab:gbr_persistence}). The subsample columns count selections in the five expanding and five rolling windows at 30 days, and in all 70 window-maturity combinations (Table~\ref{tab:gbr_window_selections}).
\end{flushleft}
\end{table}

\subsection{Descriptive Statistics and Six-Factor Alphas} \label{sec:descriptives}

Table~\ref{tab:factor_stats} reports performance statistics for the 20 option factors. Most have negative average returns over the sample. Only bakshiKurt ($t = -3.25$) and ivs\_convexity ($t = -2.57$) have average returns that differ significantly from zero at the 5\% level; ivs\_smirk is borderline ($t = -1.96$). Several factors have large negative skewness and drawdowns of more than 90\%.

\begin{table}[H]
\centering
\caption{\textbf{Performance of the Selected Option Factors}}
\label{tab:factor_stats}
\begin{small}
\begin{tabular}{lcccccc}
\toprule
\textbf{Factor} & \textbf{Ret. (\%)} & \textbf{Vol. (\%)} & \textbf{MDD (\%)} & \textbf{Sharpe} & \textbf{Skew} & \textbf{$t$-stat} \\
\midrule
bakshi\_X & -9.35 & 26.56 & -93.39 & -0.35 & 0.14 & -1.32 \\
bakshiKurt & -11.23 & 12.84 & -91.07 & -0.87 & -2.29 & -3.25 \\
BTX\_alpha\_plus & 0.56 & 22.68 & -75.07 & 0.02 & -0.31 & 0.10 \\
BTX\_EQCV\_k & 0.57 & 25.98 & -78.32 & 0.02 & 0.49 & 0.09 \\
BTX\_EQV\_relRightTail\_10 & 5.04 & 18.08 & -46.76 & 0.28 & -0.05 & 1.23 \\
BTX\_EQV\_relRightTail\_k & -0.31 & 23.91 & -71.94 & -0.01 & 0.74 & -0.05 \\
BTX\_EQV\_relTail\_k & -2.35 & 18.81 & -81.12 & -0.13 & 0.27 & -0.47 \\
BTX\_LJV\_7\_5 & -4.42 & 24.46 & -87.35 & -0.18 & 0.12 & -0.73 \\
BTX\_MLJI\_20\_annual & -1.45 & 23.74 & -80.28 & -0.06 & 0.91 & -0.26 \\
BTX\_MRJI\_20\_annual & -2.35 & 27.83 & -84.50 & -0.08 & 0.35 & -0.35 \\
BTX\_MRJI\_k\_annual & -1.12 & 27.35 & -79.96 & -0.04 & 0.30 & -0.17 \\
BTX\_RJV\_10 & 0.81 & 27.46 & -76.35 & 0.03 & 0.36 & 0.12 \\
BTX\_RJV\_k & 0.05 & 27.41 & -81.68 & 0.00 & 0.21 & 0.01 \\
ivs\_convexity & -14.78 & 17.94 & -96.94 & -0.82 & -3.59 & -2.57 \\
ivs\_smirk & -11.17 & 18.93 & -93.05 & -0.59 & -2.52 & -1.96 \\
P\_implQ\_var & -5.11 & 26.62 & -92.39 & -0.19 & -2.09 & -0.71 \\
P\_implQ\_var\_eps & -3.28 & 23.10 & -79.82 & -0.14 & 0.53 & -0.60 \\
RND\_kurt\_eps & -4.90 & 14.91 & -70.53 & -0.33 & -0.30 & -1.35 \\
RND\_mu & 8.26 & 31.16 & -67.19 & 0.27 & 0.64 & 1.06 \\
SVIX2\_put\_annual & -9.00 & 29.77 & -94.91 & -0.30 & 0.14 & -1.24 \\
\bottomrule
\end{tabular}
\end{small}
\begin{flushleft}
\footnotesize This table reports performance statistics for the 20 long--short option factors selected by the screen, February 2004 to July 2023. Ret.\ is the annualized mean return, Vol.\ the annualized standard deviation, MDD the maximum drawdown, Sharpe the annualized Sharpe ratio, and Skew the skewness of monthly returns. The $t$-statistic tests whether the mean return is zero and uses Newey--West standard errors with 6 lags.
\end{flushleft}
\end{table}

Table~\ref{tab:ff_regressions} reports regressions of the option factors on the Fama--French five factors and momentum. The six factors explain up to 59\% of the variation of an option factor (bakshi\_X). Most option factors load positively on the market and negatively on profitability (RMW) and momentum. Four annualized alphas are significant at the 5\% level, all negative: bakshiKurt ($-10.5\%$), ivs\_convexity ($-16.7\%$), ivs\_smirk ($-12.6\%$), and SVIX2\_put\_annual ($-9.1\%$). The remaining alphas are not significantly different from zero. This low-dimensional benchmark omits most of the equity factor zoo, which is why we turn to the DS-LASSO.

\begin{center}
\begin{scriptsize}
\begin{longtable}{lcccccccc}
\caption{\textbf{Six-Factor Alphas and Loadings of the Selected Option Factors}} \label{tab:ff_regressions} \\
\toprule
\textbf{Factor} & \textbf{Alpha (\%)} & \textbf{Mkt-RF} & \textbf{SMB} & \textbf{HML} & \textbf{RMW} & \textbf{CMA} & \textbf{Mom} & \textbf{Adj. $R^2$} \\
\midrule
\endfirsthead

\multicolumn{9}{c}%
{{\bfseries \tablename\ \thetable{} -- continued from previous page}} \\
\toprule
\textbf{Factor} & \textbf{Alpha (\%)} & \textbf{Mkt-RF} & \textbf{SMB} & \textbf{HML} & \textbf{RMW} & \textbf{CMA} & \textbf{Mom} & \textbf{Adj. $R^2$} \\
\midrule
\endhead

\bottomrule
\endfoot

\bottomrule
\multicolumn{9}{p{\textwidth}}{\vspace{2pt}\footnotesize This table reports time-series regressions of each option factor on the Fama--French five factors and momentum, February 2004 to July 2023 (234 months). Alphas are annualized, in percent. $t$-statistics based on Newey--West standard errors with 6 lags are reported in parentheses.} \\
\endlastfoot

bakshi\_X & -7.82 & 0.67 & -0.06 & 0.36 & -1.53 & -0.69 & -0.51 & 0.59 \\
 & (-1.87) & (6.12) & (-0.28) & (2.13) & (-6.59) & (-3.58) & (-4.51) & \\
\addlinespace
bakshiKurt & -10.53 & 0.00 & -0.31 & 0.06 & -0.13 & 0.07 & 0.22 & 0.10 \\
 & (-3.14) & (0.02) & (-1.97) & (0.60) & (-0.64) & (0.58) & (1.71) & \\
\addlinespace
BTX\_alpha\_plus & 1.47 & -0.51 & -0.37 & -0.12 & 0.80 & 0.40 & 0.39 & 0.47 \\
 & (0.38) & (-7.07) & (-2.89) & (-0.72) & (3.20) & (2.22) & (4.26) & \\
\addlinespace
BTX\_EQCV\_k & 0.59 & 0.59 & 0.18 & 0.25 & -1.10 & -0.74 & -0.45 & 0.49 \\
 & (0.13) & (7.27) & (0.95) & (1.24) & (-3.97) & (-3.62) & (-4.26) & \\
\addlinespace
BTX\_EQV\_relRightTail\_10 & 2.31 & 0.31 & 0.59 & -0.52 & -0.22 & 0.27 & -0.24 & 0.28 \\
 & (0.71) & (3.43) & (5.13) & (-2.46) & (-0.90) & (1.23) & (-3.12) & \\
\addlinespace
BTX\_EQV\_relRightTail\_k & -2.75 & 0.54 & 0.51 & 0.05 & -0.56 & -0.31 & -0.49 & 0.49 \\
 & (-0.68) & (6.80) & (3.58) & (0.37) & (-1.91) & (-1.90) & (-6.11) & \\
\addlinespace
BTX\_EQV\_relTail\_k & -2.40 & 0.35 & 0.15 & 0.15 & -0.67 & 0.08 & -0.45 & 0.45 \\
 & (-0.70) & (6.53) & (0.90) & (1.23) & (-3.52) & (0.52) & (-5.27) & \\
\addlinespace
BTX\_LJV\_7\_5 & -2.57 & 0.45 & 0.24 & 0.35 & -1.21 & -0.57 & -0.47 & 0.51 \\
 & (-0.71) & (3.86) & (1.44) & (2.55) & (-5.15) & (-2.84) & (-5.07) & \\
\addlinespace
BTX\_MLJI\_20\_annual & -1.04 & 0.53 & 0.15 & 0.29 & -1.07 & -0.27 & -0.48 & 0.51 \\
 & (-0.28) & (7.10) & (0.89) & (1.47) & (-4.32) & (-1.14) & (-5.34) & \\
\addlinespace
BTX\_MRJI\_20\_annual & -1.44 & 0.55 & 0.34 & 0.07 & -1.28 & -0.54 & -0.52 & 0.47 \\
 & (-0.32) & (3.73) & (1.76) & (0.38) & (-3.68) & (-2.53) & (-5.41) & \\
\addlinespace
BTX\_MRJI\_k\_annual & -2.58 & 0.67 & 0.47 & 0.18 & -1.00 & -0.57 & -0.50 & 0.55 \\
 & (-0.65) & (6.77) & (2.80) & (1.03) & (-3.38) & (-3.39) & (-4.99) & \\
\addlinespace
BTX\_RJV\_10 & 0.90 & 0.52 & 0.46 & -0.11 & -1.11 & -0.33 & -0.55 & 0.45 \\
 & (0.21) & (3.52) & (2.43) & (-0.62) & (-3.12) & (-1.79) & (-5.77) & \\
\addlinespace
BTX\_RJV\_k & -0.89 & 0.68 & 0.40 & 0.17 & -1.12 & -0.64 & -0.49 & 0.55 \\
 & (-0.21) & (6.20) & (2.41) & (1.01) & (-3.71) & (-3.69) & (-5.08) & \\
\addlinespace
ivs\_convexity & -16.71 & 0.38 & 0.01 & 0.07 & -0.36 & 0.19 & 0.01 & 0.12 \\
 & (-2.79) & (2.54) & (0.05) & (0.45) & (-1.51) & (1.22) & (0.11) & \\
\addlinespace
ivs\_smirk & -12.64 & 0.40 & -0.05 & 0.20 & -0.43 & -0.18 & -0.16 & 0.20 \\
 & (-2.47) & (3.22) & (-0.23) & (1.43) & (-1.67) & (-0.89) & (-2.20) & \\
\addlinespace
P\_implQ\_var & -9.63 & 0.76 & 0.33 & -0.24 & -0.60 & -0.29 & -0.38 & 0.41 \\
 & (-1.62) & (4.40) & (2.07) & (-0.92) & (-3.48) & (-0.94) & (-3.75) & \\
\addlinespace
P\_implQ\_var\_eps & -4.90 & 0.47 & 0.47 & 0.27 & -0.55 & -0.43 & -0.37 & 0.44 \\
 & (-1.20) & (4.36) & (2.89) & (1.93) & (-2.52) & (-2.00) & (-2.47) & \\
\addlinespace
RND\_kurt\_eps & -4.85 & -0.17 & 0.07 & -0.13 & 0.26 & 0.46 & 0.03 & 0.07 \\
 & (-1.35) & (-1.50) & (0.45) & (-0.64) & (1.35) & (1.99) & (0.40) & \\
\addlinespace
RND\_mu & 11.39 & -0.86 & -0.09 & -0.34 & 0.94 & 0.48 & 0.54 & 0.47 \\
 & (1.86) & (-4.27) & (-0.33) & (-1.23) & (3.29) & (1.77) & (4.02) & \\
\addlinespace
SVIX2\_put\_annual & -9.07 & 0.81 & 0.05 & 0.39 & -1.47 & -0.82 & -0.52 & 0.58 \\
 & (-2.08) & (6.25) & (0.20) & (1.87) & (-6.47) & (-3.51) & (-3.88) & \\

\end{longtable}
\end{scriptsize}
\end{center}

\subsection{Individual Tests with the DS-LASSO} \label{sec:dsresults}

Table~\ref{tab:ds_results} reports the DS-LASSO estimates for each of the 20 option factors against the 160 equity factors. Eight of the 20 have SDF loadings that differ significantly from zero at the 5\% level: bakshiKurt, BTX\_alpha\_plus, BTX\_EQV\_relTail\_k, BTX\_LJV\_7\_5, BTX\_MRJI\_20\_annual, ivs\_convexity, P\_implQ\_var, and RND\_kurt\_eps. After a Bonferroni correction for testing 20 factors, four remain: ivs\_convexity ($t = 3.92$), BTX\_EQV\_relTail\_k ($t = 3.65$), BTX\_LJV\_7\_5 ($t = 3.59$), and bakshiKurt ($t = 3.15$). The Benjamini--Hochberg procedure adds RND\_kurt\_eps ($t = 2.89$). The same four factors clear the $t > 3.0$ hurdle of \citet{harvey2016}.

\begin{table}[H]
\centering
\caption{\textbf{DS-LASSO Tests of the Option Factors}}
\label{tab:ds_results}
\begin{small}
\begin{tabular}{lccccc}
\toprule
\textbf{Factor} & \textbf{$\lambda_s$ (bp)} & \textbf{$t$-stat} & \textbf{Avg.\ ret.\ (bp)} & \textbf{Bonf.} & \textbf{BH} \\
\midrule
bakshi\_X & 65.30 & 0.98 & -77.9 &  &  \\
bakshiKurt & 78.51 & 3.15$^{***}$ & -93.6 & \checkmark & \checkmark \\
BTX\_alpha\_plus & 74.37 & 2.23$^{**}$ & 4.7 &  &  \\
BTX\_EQCV\_k & -33.84 & -0.30 & 4.7 &  &  \\
BTX\_EQV\_relRightTail\_10 & -55.49 & -0.39 & 42.0 &  &  \\
BTX\_EQV\_relRightTail\_k & -127.68 & -1.82$^{*}$ & -2.6 &  &  \\
BTX\_EQV\_relTail\_k & 295.11 & 3.65$^{***}$ & -19.6 & \checkmark & \checkmark \\
BTX\_LJV\_7\_5 & 445.65 & 3.59$^{***}$ & -36.8 & \checkmark & \checkmark \\
BTX\_MLJI\_20\_annual & -170.84 & -1.93$^{*}$ & -12.1 &  &  \\
BTX\_MRJI\_20\_annual & -310.38 & -2.28$^{**}$ & -19.6 &  &  \\
BTX\_MRJI\_k\_annual & 66.16 & 0.79 & -9.3 &  &  \\
BTX\_RJV\_10 & 43.87 & 1.76$^{*}$ & 6.8 &  &  \\
BTX\_RJV\_k & 115.91 & 1.79$^{*}$ & 0.4 &  &  \\
ivs\_convexity & 420.03 & 3.92$^{***}$ & -123.2 & \checkmark & \checkmark \\
ivs\_smirk & -45.34 & -0.88 & -93.1 &  &  \\
P\_implQ\_var & -260.97 & -2.40$^{**}$ & -42.6 &  &  \\
P\_implQ\_var\_eps & 62.55 & 0.69 & -27.3 &  &  \\
RND\_kurt\_eps & 150.55 & 2.89$^{***}$ & -40.8 &  & \checkmark \\
RND\_mu & -140.93 & -1.32 & 68.8 &  &  \\
SVIX2\_put\_annual & 59.72 & 0.40 & -75.0 &  &  \\
\bottomrule
\end{tabular}
\end{small}
\begin{flushleft}
\footnotesize This table reports DS-LASSO tests of each option factor against the 160 equity factors, following \citet{feng2020taming}, February 2004 to July 2023. Each factor is tested individually. $\lambda_s$ is the SDF loading scaled to a unit univariate beta exposure, in basis points per month. The $t$-statistic is based on the heteroskedasticity-robust asymptotic variance of \citet{feng2020taming}. Avg.\ ret.\ is the factor's average return in basis points per month (the annualized mean in Table~\ref{tab:factor_stats} divided by 12). $^{*}$, $^{**}$, and $^{***}$ denote significance at the 10\%, 5\%, and 1\% levels. Bonf.\ marks factors that remain significant after a Bonferroni correction for testing 20 factors ($|t| > 3.02$), and BH marks factors that remain significant under the \citet{benjamini1995controlling} procedure with a false discovery rate of 5\%.
\end{flushleft}
\end{table}

The signs of the significant loadings follow a pattern. Measures of tail size and curvature have positive loadings: risk-neutral kurtosis (bakshiKurt) and its idiosyncratic component (RND\_kurt\_eps), implied-volatility convexity (ivs\_convexity), left-tail jump variation (BTX\_LJV\_7\_5), and the share of risk-neutral variance due to large jumps in either tail (BTX\_EQV\_relTail\_k). Right-tail jump intensity (BTX\_MRJI\_20\_annual) has a negative loading, and the right-tail decay parameter (BTX\_alpha\_plus), which is high when large upward jumps are unlikely, has a positive one. Option-implied physical variance (P\_implQ\_var) has a negative loading. These directions are consistent with investors disliking exposure to tail and downside risk \citep{bollerslev2015efficient}, being willing to pay for right-skewed payoffs \citep{conrad2013ex}, and avoiding high-volatility stocks \citep{ang2006cross}. Our design, however, does not identify the economic mechanism.

For five of the eight significant factors, the sign of $\lambda_s$ differs from the sign of the factor's average return: bakshiKurt, RND\_kurt\_eps, ivs\_convexity, BTX\_LJV\_7\_5, and BTX\_EQV\_relTail\_k have positive loadings but negative average returns. As explained in Section~\ref{sec:dslasso}, $\lambda_s$ is the price of exposure to a factor conditional on the selected equity controls, whereas the average return also reflects the factor's exposure to those controls. The option factors load heavily on equity factors with large average returns, in particular negatively on RMW and momentum (Table~\ref{tab:ff_regressions}), and such exposures can reverse the sign. For bakshiKurt and ivs\_convexity, the six-factor loadings are small and the alphas significantly negative, so the gap cannot be traced to these six exposures. In addition, $\lambda_s$ is estimated from the cross-section of test portfolios, including the $3 \times 2$ portfolios sorted on each option characteristic and size, rather than from the time series of the long--short factor. We therefore read the sign of $\lambda_s$ as a statement about which direction of covariance with the factor is priced in the cross-section of test assets, not as the premium earned by the long--short portfolio.

\subsection{Aggregate Tests} \label{sec:aggregate}

\noindent\textbf{Joint alphas and the maximum Sharpe ratio.} The GRS test of whether the alphas of the 20 option factors relative to the 160 equity factors are jointly zero yields a statistic of 1.25 with a $p$-value of 0.25. The test does not reject. With 20 test assets, 160 factors, and 234 months, the test has few residual degrees of freedom and limited power, so this non-rejection is weak evidence in either direction. In terms of Sharpe ratios, the 160 equity factors reach an in-sample annualized maximum Sharpe ratio of 3.35, the 20 option factors alone 1.29, and the combined set 3.74 (Appendix Table~\ref{tab:spanning_robustness}). Given the upward bias of in-sample Sharpe ratios in large sets of assets, the level of 3.35 is an upper bound rather than an achievable Sharpe ratio. The GRS result shows that the increase of 0.39 is not statistically significant. The option factors therefore do not measurably expand the investment opportunity set, even though some of them carry non-zero SDF loadings individually. This is consistent with a small number of relevant factors being diluted in a joint test of 20.\\

\noindent\textbf{Shared variation.} Table~\ref{tab:cca_results} reports the redundancy indices. The 160 equity factors explain 91.0\% of the variance of the 20 option factors, well above the noise level of about 69\%. Only 9.0\% of the option factors' variance is unique to them. In the other direction, the 20 option factors explain 35.4\% of the variance of the equity factors, far above the noise level of about 9\% but leaving 64.6\% unique to the equity factors. Option factors are thus largely spanned by the equity factor zoo in terms of variance, while they capture about a third of the variation in the equity factors.

\begin{table}[H]
\centering
\begin{threeparttable}
\caption{\textbf{Redundancy between Option and Equity Factors}}
\label{tab:cca_results}
\begin{small}
\begin{tabular}{lcc}
\toprule
\textbf{Measure} & \textbf{Value} & \textbf{Noise level} \\
\midrule
Options explained by equity factors ($RI_{O|E}$) & 0.9103 & 0.69 \\
Equity factors explained by options ($RI_{E|O}$) & 0.3536 & 0.09 \\
\midrule
Unique share of options ($1 - RI_{O|E}$) & 0.0897 & \\
Unique share of equity factors ($1 - RI_{E|O}$) & 0.6464 & \\
\bottomrule
\end{tabular}
\end{small}
\begin{tablenotes}
\footnotesize \item This table reports Stewart--Love redundancy indices from a canonical correlation analysis of the 20 selected option factors ($O$) and the 160 equity factors ($E$), estimated on standardized monthly returns with all 20 canonical pairs, February 2004 to July 2023. $RI_{Y|X}$ is the average share of the variance of the factors in $Y$ that is explained in-sample by linear combinations of the factors in $X$. The noise level is the approximate value of the index for independent noise of the same dimensions, $p_X/(T-1)$, where $p_X$ is the number of factors in $X$ and $T = 234$.
\end{tablenotes}
\end{threeparttable}
\end{table}

\subsection{Robustness} \label{sec:robustness}

\noindent\textbf{Option maturity.} We rerun the entire analysis, including the screen, for option-implied characteristics at maturities of 1, 2, 3, 5, 10, 30, and 60 days (Appendix~C). The aggregate results are stable. The equity factors explain between 91\% and 94\% of the option factors' variance at every maturity, and the option factors explain between 29\% and 35\% of the equity factors' variance (Table~\ref{tab:cca_robustness}). The GRS test does not reject at the 5\% level at six of the seven maturities; at 60 days, $p = 0.043$ (Table~\ref{tab:grs_robustness}). With seven maturities, one rejection at the 5\% level is not unusual even if the null is true everywhere. The in-sample Sharpe ratio gain ranges from 0.16 to 0.39 (Table~\ref{tab:spanning_robustness}).

The factors selected by the screen, and therefore the individual tests, differ across maturities (Tables~\ref{tab:gbr_persistence} and \ref{tab:ds_lasso_fred}). Some results are robust. Left-tail jump variation (BTX\_LJV\_7\_5) is selected at five maturities and has a positive and significant loading at each of them. The share of variance from large jumps (BTX\_EQV\_relTail\_k) has positive and significant loadings at all three maturities at which it is selected. Implied-volatility convexity has positive and significant loadings at 1, 2, and 30 days, but a negative loading at 10 days. Other results appear only at 30 days, notably those for the kurtosis measures bakshiKurt and RND\_kurt\_eps and for BTX\_MRJI\_20\_annual. The loading on option-implied physical variance changes sign across maturities.\\

\noindent\textbf{Risk-free rate.} Replacing the FRED Treasury-bill rate with the option-implied TOPT rate leaves the results largely unchanged (Tables~\ref{tab:ds_lasso_fred} and \ref{tab:ds_lasso_topt}). The rate cancels out of the long--short factors, and a common shift in the test assets' excess returns is absorbed by the intercept $\gamma_0$. Differences therefore arise mainly because the LASSO steps select slightly different controls. At 30 days, no factor changes significance at the 5\% level. At other maturities a few results differ, for example bakshiImplVar at 10 days ($t = 2.65$ with FRED and $0.59$ with TOPT) and BTX\_RJV\_k at 2 days ($t = -1.26$ and $-2.52$).

\section{Discussion and Limitations} \label{sec:discussion}

The three sets of results fit together. In terms of variance, option factors are largely spanned by the equity factor zoo: the equity factors reproduce about 91\% of their monthly variation, and as a group they neither produce jointly significant alphas nor significantly raise the maximum Sharpe ratio. This is not surprising, since both sets of factors are portfolios of the same stocks and many option-implied characteristics are correlated with equity characteristics such as volatility, beta, and past returns. The remaining variation is small but not irrelevant for pricing. A few option factors carry SDF loadings that the equity factors cannot replace, and the most robust of these measure the size of the tails of the risk-neutral distribution. That these factors do not show up in the joint tests is consistent with their being a small subset of the 20 factors tested. Option prices thus appear to add information about tail risk that characteristics built from past returns and accounting data do not fully capture, while most other option-implied information is already reflected in the equity factor zoo.

Our design has several limitations. First, all results are in-sample. The screen and the tests use the same observations, so the test results are conditional on the screen, and we do not evaluate the selected factors on a hold-out period. With 234 months, splitting the sample would substantially reduce power; we leave out-of-sample validation to future work. Second, the 20 factors were selected from 137, so the effective number of tests exceeds 20, and even the Bonferroni correction understates the multiple-testing problem. Third, the selection of individual factors varies across maturities and subsamples. Finally, the in-sample maximum Sharpe ratios are biased upward, and we do not report out-of-sample or constrained portfolios.

\section{Conclusion} \label{sec:conclusion}

We ask whether factors sorted on option-implied characteristics add pricing information to the equity factor zoo. Using 137 option-implied characteristics, a screen to 20 representative factors, and the DS-LASSO test of \citet{feng2020taming} against 160 equity factors, we find that option factors are largely, but not entirely, spanned. The equity factors explain about 91\% of the option factors' variance in-sample, and adding the option factors does not significantly improve the maximum Sharpe ratio. Individually, eight option factors have non-zero SDF loadings at the 5\% level, and four remain significant after a Bonferroni correction: implied-volatility convexity, the share of variance due to large jumps, left-tail jump variation, and risk-neutral kurtosis. The most robust results across option maturities concern the jump-tail measures.

Two extensions seem most useful. First, an out-of-sample design, for example screening on one part of the sample and testing on another, would remove the dependence of our tests on the screen. Second, extending the analysis beyond S\&P 500 constituents, and studying whether the incremental information is concentrated in particular market conditions, would show how general the tail-risk dimension is.

\newpage
\bibliography{2_references}
\newpage
\begin{appendices}

\section*{Appendix A: Equity Characteristics}
\label{sec:appendix_a}
\setcounter{table}{0}
\renewcommand{\thetable}{A.\arabic{table}}

Table~\ref{tab:appendix_characteristics} lists the equity characteristics from \citet{jensen2023replication} from which we construct the equity factors. We use 152 of the 153 characteristics. Price per share (prc, No.~125) is excluded because it has no clear economic interpretation as a sorting variable.

{\scriptsize
\begin{longtable}{>{\raggedright\arraybackslash}p{0.8cm}
                  >{\raggedright\arraybackslash}p{3.5cm}
                  >{\raggedright\arraybackslash}p{7.5cm}
                  >{\raggedright\arraybackslash}p{4.5cm}}
\caption{\textbf{Equity Characteristics}} \label{tab:appendix_characteristics} \\

\toprule
No. & Abbreviation & Description & Authors \\
\midrule
\endfirsthead

\toprule
No. & Abbreviation & Description & Authors \\
\midrule
\endhead

\midrule
\endfoot

\bottomrule
\endlastfoot

1 & niq\_su & Standardized earnings surprise & Foster, Olsen, and Shevlin (1984) \\
2 & ret\_6\_1 & Price momentum t--6 to t--1 & Jegadeesh and Titman (1993) \\
3 & ret\_12\_1 & Price momentum t--12 to t--1 & Jegadeesh and Titman (1993) \\
4 & saleq\_su & Standardized revenue surprise & Jegadeesh and Livnat (2006) \\
5 & tax\_gr1a & Tax expense surprise & Thomas and Zhang (2011) \\
6 & ni\_inc8q & Consecutive quarters with earnings increases & Barth, Elliott, and Finn (1999) \\
7 & prc\_highprc\_252d & Price relative to yearly high & George and Hwang (2004) \\
8 & resff3\_6\_1 & Residual momentum t--6 to t--1 & Blitz, Huij, and Martens (2011) \\
9 & resff3\_12\_1 & Residual momentum t--12 to t--1 & Blitz, Huij, and Martens (2011) \\
10 & be\_me & Book-to-market equity & Rosenberg, Reid, and Lanstein (1985) \\
11 & debt\_me & Debt-to-market & Bhandari (1988) \\
12 & at\_me & Assets-to-market & Fama and French (1992) \\
13 & ret\_60\_12 & Long-term reversal & De Bondt and Thaler (1985) \\
14 & ni\_me & Earnings-to-price & Basu (1983) \\
15 & fcf\_me & Free cash flow-to-price & Lakonishok, Shleifer, and Vishny (1994) \\
16 & div12m\_me & Dividend yield & Litzenberger and Ramaswamy (1979) \\
17 & eqpo\_me & Payout yield & Boudoukh et al. (2007) \\
18 & eqnpo\_me & Net payout yield & Boudoukh et al. (2007) \\
19 & sale\_gr3 & Sales growth (3 years) & Lakonishok, Shleifer, and Vishny (1994) \\
20 & sale\_gr1 & Sales growth (1 year) & Lakonishok, Shleifer, and Vishny (1994) \\
21 & ebitda\_mev & EBITDA-to-market enterprise value & Loughran and Wellman (2011) \\
22 & sale\_me & Sales-to-market & Barbee, Mukherji, and Raines (1996) \\
23 & ocf\_me & Operating cash flow-to-market & Desai, Rajgopal, and Venkatachalam (2004) \\
24 & ival\_me & Intrinsic value-to-market & Frankel and Lee (1998) \\
25 & bev\_mev & Book-to-market enterprise value & Penman, Richardson, and Tuna (2007) \\
26 & netdebt\_me & Net debt-to-price & Penman, Richardson, and Tuna (2007) \\
27 & eq\_dur & Equity duration & Dechow, Sloan, and Soliman (2004) \\
28 & capex\_abn & Abnormal corporate investment & Titman, Wei, and Xie (2004) \\
29 & at\_gr1 & Asset growth & Cooper, Gulen, and Schill (2008) \\
30 & ppeinv\_gr1a & Change in PPE and inventory & Lyandres, Sun, and Zhang (2008) \\
31 & noa\_at & Net operating assets & Hirshleifer et al. (2004) \\
32 & noa\_gr1a & Change in net operating assets & Hirshleifer et al. (2004) \\
33 & lnoa\_gr1a & Change in long-term net operating assets & Fairfield, Whisenant, and Yohn (2003) \\
34 & capx\_gr1 & CAPEX growth (1 year) & Xie (2001) \\
35 & capx\_gr2 & CAPEX growth (2 years) & Anderson and Garcia-Feijoo (2006) \\
36 & capx\_gr3 & CAPEX growth (3 years) & Anderson and Garcia-Feijoo (2006) \\
37 & chcsho\_12m & Net stock issues & Pontiff and Woodgate (2008) \\
38 & eqnpo\_12m & Equity net payout & Daniel and Titman (2006) \\
39 & debt\_gr3 & Growth in book debt (3 years) & Lyandres, Sun, and Zhang (2008) \\
40 & inv\_gr1 & Inventory growth & Belo and Lin (2011) \\
41 & inv\_gr1a & Inventory change & Thomas and Zhang (2002) \\
42 & oaccruals\_at & Operating accruals & Sloan (1996) \\
43 & taccruals\_at & Total accruals & Richardson et al. (2005) \\
44 & cowc\_gr1a & Change in current operating working capital & Richardson et al. (2005) \\
45 & coa\_gr1a & Change in current operating assets & Richardson et al. (2005) \\
46 & col\_gr1a & Change in current operating liabilities & Richardson et al. (2005) \\
47 & nncoa\_gr1a & Change in net noncurrent operating assets & Richardson et al. (2005) \\
48 & ncoa\_gr1a & Change in noncurrent operating assets & Richardson et al. (2005) \\
49 & ncol\_gr1a & Change in noncurrent operating liabilities & Richardson et al. (2005) \\
50 & nfna\_gr1a & Change in net financial assets & Richardson et al. (2005) \\
51 & sti\_gr1a & Change in short-term investments & Richardson et al. (2005) \\
52 & lti\_gr1a & Change in long-term investments & Richardson et al. (2005) \\
53 & fnl\_gr1a & Change in financial liabilities & Richardson et al. (2005) \\
54 & be\_gr1a & Change in common equity & Richardson et al. (2005) \\
55 & oaccruals\_ni & Percent operating accruals & Hafzalla, Lundholm, and Van Winkle (2011) \\
56 & taccruals\_ni & Percent total accruals & Hafzalla, Lundholm, and Van Winkle (2011) \\
57 & netis\_at & Net total issuance & Bradshaw, Richardson, and Sloan (2006) \\
58 & eqnetis\_at & Net equity issuance & Bradshaw, Richardson, and Sloan (2006) \\
59 & dbnetis\_at & Net debt issuance & Bradshaw, Richardson, and Sloan (2006) \\
60 & niq\_be & Quarterly return on equity & Hou, Xue, and Zhang (2015) \\
61 & niq\_be\_chg1 & Change in quarterly return on equity & Hou, Xue, and Zhang (2015) \\
62 & niq\_at & Quarterly return on assets & Balakrishnan, Bartov, and Faurel (2010) \\
63 & niq\_at\_chg1 & Change in quarterly return on assets & Balakrishnan, Bartov, and Faurel (2010) \\
64 & ebit\_bev & Return on net operating assets & Soliman (2008) \\
65 & ebit\_sale & Profit margin & Soliman (2008) \\
66 & sale\_bev & Asset turnover & Soliman (2008) \\
67 & at\_turnover & Capital turnover & Haugen and Baker (1996) \\
68 & gp\_at & Gross profits-to-assets & Novy-Marx (2013) \\
69 & gp\_atl1 & Gross profits-to-lagged assets & Novy-Marx (2013) \\
70 & ope\_be & Operating profits-to-book equity & Fama and French (2015) \\
71 & ope\_bel1 & Operating profits-to-lagged book equity & Fama and French (2015) \\
72 & op\_at & Operating profits-to-book assets & Ball et al. (2016) \\
73 & op\_atl1 & Operating profits-to-lagged book assets & Ball et al. (2016) \\
74 & cop\_at & Cash-based operating profits-to-book assets & Ball et al. (2016) \\
75 & cop\_atl1 & Cash-based operating profits-to-lagged book assets & Ball et al. (2016) \\
76 & f\_score & Piotroski F-score & Piotroski (2000) \\
77 & o\_score & Ohlson O-score & Dichev (1998) \\
78 & z\_score & Altman Z-score & Dichev (1998) \\
79 & pi\_nix & Taxable income-to-book income & Lev and Nissim (2004) \\
80 & at\_be & Book leverage & Fama and French (1992) \\
81 & saleq\_gr1 & Sales growth (1 quarter) & Lakonishok, Shleifer, and Vishny (1994) \\
82 & rd\_me & R\&D-to-market & Chan, Lakonishok, and Sougiannis (2001) \\
83 & rd\_sale & R\&D-to-sales & Chan, Lakonishok, and Sougiannis (2001) \\
84 & opex\_at & Operating leverage & Novy-Marx (2011) \\
85 & emp\_gr1 & Hiring rate & Belo, Lin, and Bazdresch (2014) \\
86 & rd5\_at & R\&D capital-to-book assets & Li (2011) \\
87 & age & Firm age & Jiang, Lee, and Zhang (2005) \\
88 & dsale\_dinv & Change in sales minus change in inventory & Abarbanell and Bushee (1998) \\
89 & dsale\_drec & Change in sales minus change in receivables & Abarbanell and Bushee (1998) \\
90 & dgp\_dsale & Change in gross margin minus change in sales & Abarbanell and Bushee (1998) \\
91 & dsale\_dsga & Change in sales minus change in SG\&A & Abarbanell and Bushee (1998) \\
92 & sale\_emp\_gr1 & Labor force efficiency & Abarbanell and Bushee (1998) \\
93 & tangibility & Asset tangibility & Hahn and Lee (2009) \\
94 & kz\_index & Kaplan--Zingales index & Lamont, Polk, and Saa-Requejo (2001) \\
95 & ocfq\_saleq\_std & Cash flow volatility & Huang (2009) \\
96 & cash\_at & Cash-to-assets & Palazzo (2012) \\
97 & ni\_ar1 & Earnings persistence & Francis et al. (2004) \\
98 & ni\_ivol & Earnings volatility & Francis et al. (2004) \\
99 & earnings\_variability & Earnings variability & Francis et al. (2004) \\
100 & aliq\_at & Liquidity of book assets & Ortiz-Molina and Phillips (2014) \\
101 & aliq\_mat & Liquidity of market assets & Ortiz-Molina and Phillips (2014) \\
102 & seas\_1\_1an & Year 1-lagged return, annual & Heston and Sadka (2008) \\
103 & seas\_1\_1na & Year 1-lagged return, nonannual & Heston and Sadka (2008) \\
104 & seas\_2\_5an & Years 2--5 lagged returns, annual & Heston and Sadka (2008) \\
105 & seas\_2\_5na & Years 2--5 lagged returns, nonannual & Heston and Sadka (2008) \\
106 & seas\_6\_10an & Years 6--10 lagged returns, annual & Heston and Sadka (2008) \\
107 & seas\_6\_10na & Years 6--10 lagged returns, nonannual & Heston and Sadka (2008) \\
108 & seas\_11\_15an & Years 11--15 lagged returns, annual & Heston and Sadka (2008) \\
109 & seas\_11\_15na & Years 11--15 lagged returns, nonannual & Heston and Sadka (2008) \\
110 & seas\_16\_20an & Years 16--20 lagged returns, annual & Heston and Sadka (2008) \\
111 & seas\_16\_20na & Years 16--20 lagged returns, nonannual & Heston and Sadka (2008) \\
112 & market\_equity & Market equity & Banz (1981) \\
113 & ivol\_ff3\_21d & Idiosyncratic volatility (FF3) & Ang et al. (2006) \\
114 & ivol\_capm\_252d & Idiosyncratic volatility (CAPM, 252 days) & Ali, Hwang, and Trombley (2003) \\
115 & ivol\_capm\_21d & Idiosyncratic volatility (CAPM, 21 days) & Ang et al. (2006) \\
116 & ivol\_hxz4\_21d & Idiosyncratic volatility (q-factor) & Hou, Xue, and Zhang (2015) \\
117 & rvol\_21d & Return volatility & Ang et al. (2006) \\
118 & beta\_60m & Market beta & Fama and MacBeth (1973) \\
119 & betabab\_1260d & Frazzini--Pedersen market beta & Frazzini and Pedersen (2014) \\
120 & beta\_dimson\_21d & Dimson beta & Dimson (1979) \\
121 & turnover\_126d & Share turnover & Datar, Naik, and Radcliffe (1998) \\
122 & turnover\_var\_126d & CV of share turnover & Chordia, Subrahmanyam, and Anshuman (2001) \\
123 & dolvol\_126d & Dollar trading volume & Brennan, Chordia, and Subrahmanyam (1998) \\
124 & dolvol\_var\_126d & CV of dollar trading volume & Chordia, Subrahmanyam, and Anshuman (2001) \\
125 & prc & Price per share (not used) & Miller and Scholes (1982) \\
126 & ami\_126d & Amihud illiquidity measure & Amihud (2002) \\
127 & zero\_trades\_21d & Zero-trade days (1 month) & Liu (2006) \\
128 & zero\_trades\_126d & Zero-trade days (6 months) & Liu (2006) \\
129 & zero\_trades\_252d & Zero-trade days (12 months) & Liu (2006) \\
130 & rmax1\_21d & Maximum daily return & Bali, Cakici, and Whitelaw (2011) \\
131 & rskew\_21d & Total skewness & Bali, Engle, and Murray (2016) \\
132 & iskew\_capm\_21d & Idiosyncratic skewness (CAPM) & Bali, Engle, and Murray (2016) \\
133 & iskew\_ff3\_21d & Idiosyncratic skewness (FF3) & Bali, Engle, and Murray (2016) \\
134 & iskew\_hxz4\_21d & Idiosyncratic skewness (q-factor) & Hou, Xue, and Zhang (2015) \\
135 & coskew\_21d & Coskewness & Harvey and Siddique (2000) \\
136 & ret\_1\_0 & Short-term reversal & Jegadeesh (1990) \\
137 & betadown\_252d & Downside beta & Ang, Chen, and Xing (2006) \\
138 & bidaskhl\_21d & High--low bid--ask spread & Corwin and Schultz (2012) \\
139 & ret\_3\_1 & Price momentum t--3 to t--1 & Jegadeesh and Titman (1993) \\
140 & ret\_9\_1 & Price momentum t--9 to t--1 & Jegadeesh and Titman (1993) \\
141 & ret\_12\_7 & Price momentum t--12 to t--7 & Novy-Marx (2012) \\
142 & corr\_1260d & Market correlation & Asness et al. (2020) \\
143 & rmax5\_21d & Highest 5 daily returns & Bali et al. (2017) \\
144 & rmax5\_rvol\_21d & Highest 5 returns scaled by volatility & Asness et al. (2020) \\
145 & ni\_be & Return on equity & Haugen and Baker (1996) \\
146 & ocf\_at & Operating cash flow to assets & Bouchard et al. (2019) \\
147 & ocf\_at\_chg1 & Change in operating cash flow to assets & Bouchard et al. (2019) \\
148 & mispricing\_perf & Mispricing factor: Performance & Stambaugh and Yuan (2017) \\
149 & mispricing\_mgmt & Mispricing factor: Management & Stambaugh and Yuan (2017) \\
150 & qmj & Quality minus Junk (composite) & Asness, Frazzini, and Pedersen (2018) \\
151 & qmj\_prof & Quality minus Junk: Profitability & Asness, Frazzini, and Pedersen (2018) \\
152 & qmj\_growth & Quality minus Junk: Growth & Asness, Frazzini, and Pedersen (2018) \\
153 & qmj\_safety & Quality minus Junk: Safety & Asness, Frazzini, and Pedersen (2018) \\

\end{longtable}
}

\clearpage

\section*{Appendix B: Option-Implied Characteristics}
\label{sec:appendix_b}
\setcounter{table}{0}
\renewcommand{\thetable}{B.\arabic{table}}

Table~\ref{tab:appendix_opt_characteristics} lists the option-implied characteristics. The analysis uses 137 of the 138 entries. The volatility-scaled jump cutoff $k_t$ (BTX\_k, No.~60) is an input to other measures rather than a signal and is excluded. QV denotes quadratic variation under the risk-neutral measure. For the measures of \citet{bollerslev2015efficient}, the suffixes 7\_5, 10, and 20 denote log-jump cutoffs of about 7.5\%, 10\%, and 20\%, and the suffix k denotes the time-varying cutoff $k_t$, which scales with normalized at-the-money implied volatility. The suffixes \_sys and \_eps denote systematic and idiosyncratic components, and \_annual denotes annualized values. Measures without a reference are constructed in this paper.

{\scriptsize
\begin{longtable}{>{\raggedright\arraybackslash}p{0.3cm}
                  >{\raggedright\arraybackslash}p{4.4cm}
                  >{\raggedright\arraybackslash}p{7.5cm}
                  >{\raggedright\arraybackslash}p{4.5cm}}
\caption{\textbf{Option-Implied Characteristics}} \label{tab:appendix_opt_characteristics} \\

\toprule
No. & Abbreviation & Description & Authors \\
\midrule
\endfirsthead

\toprule
No. & Abbreviation & Description & Authors \\
\midrule
\endhead

\midrule
\endfoot

\bottomrule
\endlastfoot

1 & bakshi\_mu & mu (eq.~39) & Bakshi, Kapadia \& Madan (2003) \\
2 & bakshi\_V & V (eq.~7) & Bakshi, Kapadia \& Madan (2003) \\
3 & bakshi\_V\_call & V call & Bakshi, Kapadia \& Madan (2003) \\
4 & bakshi\_V\_put & V put & Bakshi, Kapadia \& Madan (2003) \\
5 & bakshi\_W & W (eq.~8) & Bakshi, Kapadia \& Madan (2003) \\
6 & bakshi\_W\_call & W call & Bakshi, Kapadia \& Madan (2003) \\
7 & bakshi\_W\_put & W put & Bakshi, Kapadia \& Madan (2003) \\
8 & bakshi\_X & X (eq.~9) & Bakshi, Kapadia \& Madan (2003) \\
9 & bakshi\_X\_call & X call & Bakshi, Kapadia \& Madan (2003) \\
10 & bakshi\_X\_put & X put & Bakshi, Kapadia \& Madan (2003) \\
11 & bakshiImplVar & Implied Variance & Bakshi, Kapadia \& Madan (2003) \\
12 & bakshiImplVar\_asym & Option-implied variance asymmetry & Huang \& Li (2019) \\
13 & bakshiImplVar\_call & Implied variance call (up) & Bakshi, Kapadia \& Madan (2003) \\
14 & bakshiImplVar\_put & Implied variance down (put) & Bakshi, Kapadia \& Madan (2003) \\
15 & bakshiImplVol\_annual & Implied volatility (annualized) & Bakshi, Kapadia \& Madan (2003) \\
16 & bakshiKurt & Kurtosis (eq.~6) & Bakshi, Kapadia \& Madan (2003) \\
17 & bakshiSkew & Skewness (eq.~5) & Bakshi, Kapadia \& Madan (2003) \\
18 & BJN\_implVol\_annual & Implied volatility (annualized) & Britten-Jones \& Neuberger (2000) \\
19 & BJN\_V\_annual & Theoretical underpinning of the VIX (annualized) & Britten-Jones \& Neuberger (2000) \\
20 & BJN\_V\_call\_annual & VIX theoretical call component (annualized) & Britten-Jones \& Neuberger (2000) \\
21 & BJN\_V\_put\_annual & VIX theoretical put component (annualized) & Britten-Jones \& Neuberger (2000) \\
22 & BTX\_alpha\_minus & Left-tail decay parameter $\alpha_t^-$ (eq.~3.3) & Bollerslev, Todorov \& Xu (2015) \\
23 & BTX\_alpha\_plus & Right-tail decay parameter $\alpha_t^+$ (eq.~3.3) & Bollerslev, Todorov \& Xu (2015) \\
24 & BTX\_EQCV\_10 & Expected continuous + small-jump QV (10\% cutoff) & Bollerslev, Todorov \& Xu (2015) \\
25 & BTX\_EQCV\_20 & Expected continuous + small-jump QV (20\% cutoff) & Bollerslev, Todorov \& Xu (2015) \\
26 & BTX\_EQCV\_7\_5 & Expected continuous + small-jump QV (7.5\% cutoff) & Bollerslev, Todorov \& Xu (2015) \\
27 & BTX\_EQCV\_k & Expected continuous + small-jump QV ($k_t$ cutoff) & Bollerslev, Todorov \& Xu (2015) \\
28 & BTX\_EQJV\_10 & Expected QV from large jumps (10\%) & Bollerslev, Todorov \& Xu (2015) \\
29 & BTX\_EQJV\_20 & Expected QV from large jumps (20\%) & Bollerslev, Todorov \& Xu (2015) \\
30 & BTX\_EQJV\_7\_5 & Expected QV from large jumps (7.5\%) & Bollerslev, Todorov \& Xu (2015) \\
31 & BTX\_EQJV\_k & Expected QV from large jumps ($k_t$) & Bollerslev, Todorov \& Xu (2015) \\
32 & BTX\_EQV\_relLeftTail\_10 & Fraction of QV from left-tail jumps (10\%) & Bollerslev, Todorov \& Xu (2015) \\
33 & BTX\_EQV\_relLeftTail\_20 & Fraction of QV from left-tail jumps (20\%) & Bollerslev, Todorov \& Xu (2015) \\
34 & BTX\_EQV\_relLeftTail\_7\_5 & Fraction of QV from left-tail jumps (7.5\%) & Bollerslev, Todorov \& Xu (2015) \\
35 & BTX\_EQV\_relLeftTail\_k & Fraction of QV from left-tail jumps ($k_t$) & Bollerslev, Todorov \& Xu (2015) \\
36 & BTX\_EQV\_relRightTail\_10 & Fraction of QV from right-tail jumps (10\%) & Bollerslev, Todorov \& Xu (2015) \\
37 & BTX\_EQV\_relRightTail\_20 & Fraction of QV from right-tail jumps (20\%) & Bollerslev, Todorov \& Xu (2015) \\
38 & BTX\_EQV\_relRightTail\_7\_5 & Fraction of QV from right-tail jumps (7.5\%) & Bollerslev, Todorov \& Xu (2015) \\
39 & BTX\_EQV\_relRightTail\_k & Fraction of QV from right-tail jumps ($k_t$) & Bollerslev, Todorov \& Xu (2015) \\
40 & BTX\_EQV\_relTail\_10 & Fraction of QV from both tails (10\%) & Bollerslev, Todorov \& Xu (2015) \\
41 & BTX\_EQV\_relTail\_20 & Fraction of QV from both tails (20\%) & Bollerslev, Todorov \& Xu (2015) \\
42 & BTX\_EQV\_relTail\_7\_5 & Fraction of QV from both tails (7.5\%) & Bollerslev, Todorov \& Xu (2015) \\
43 & BTX\_EQV\_relTail\_k & Fraction of QV from both tails ($k_t$) & Bollerslev, Todorov \& Xu (2015) \\
44 & BTX\_implContVol\_10\_annual & Implied continuous volatility (10\%) & Bollerslev, Todorov \& Xu (2015) \\
45 & BTX\_implContVol\_20\_annual & Implied continuous volatility (20\%) & Bollerslev, Todorov \& Xu (2015) \\
46 & BTX\_implContVol\_7\_5\_annual & Implied continuous volatility (7.5\%) & Bollerslev, Todorov \& Xu (2015) \\
47 & BTX\_implContVol\_k\_annual & Implied continuous volatility ($k_t$) & Bollerslev, Todorov \& Xu (2015) \\
48 & BTX\_implCrashFearVol\_10\_annual & Crash-fear volatility (10\%) & Bollerslev, Todorov \& Xu (2015) \\
49 & BTX\_implCrashFearVol\_20\_annual & Crash-fear volatility (20\%) & Bollerslev, Todorov \& Xu (2015) \\
50 & BTX\_implCrashFearVol\_7\_5\_annual & Crash-fear volatility (7.5\%) & Bollerslev, Todorov \& Xu (2015) \\
51 & BTX\_implCrashFearVol\_k\_annual & Crash-fear volatility ($k_t$) & Bollerslev, Todorov \& Xu (2015) \\
52 & BTX\_implEuphoriaVol\_10\_annual & Euphoria volatility (10\%) & Bollerslev, Todorov \& Xu (2015) \\
53 & BTX\_implEuphoriaVol\_20\_annual & Euphoria volatility (20\%) & Bollerslev, Todorov \& Xu (2015) \\
54 & BTX\_implEuphoriaVol\_7\_5\_annual & Euphoria volatility (7.5\%) & Bollerslev, Todorov \& Xu (2015) \\
55 & BTX\_implEuphoriaVol\_k\_annual & Euphoria volatility ($k_t$) & Bollerslev, Todorov \& Xu (2015) \\
56 & BTX\_implTailVol\_10\_annual & Jump volatility (10\%) & Bollerslev, Todorov \& Xu (2015) \\
57 & BTX\_implTailVol\_20\_annual & Jump volatility (20\%) & Bollerslev, Todorov \& Xu (2015) \\
58 & BTX\_implTailVol\_7\_5\_annual & Jump volatility (7.5\%) & Bollerslev, Todorov \& Xu (2015) \\
59 & BTX\_implTailVol\_k\_annual & Jump volatility ($k_t$) & Bollerslev, Todorov \& Xu (2015) \\
60 & BTX\_k & Volatility-scaled jump cutoff $k_t$ (not used) & Bollerslev, Todorov \& Xu (2015) \\
61 & BTX\_LJV\_10 & Left-tail jump variation (10\%) & Bollerslev, Todorov \& Xu (2015) \\
62 & BTX\_LJV\_20 & Left-tail jump variation (20\%) & Bollerslev, Todorov \& Xu (2015) \\
63 & BTX\_LJV\_7\_5 & Left-tail jump variation (7.5\%) & Bollerslev, Todorov \& Xu (2015) \\
64 & BTX\_LJV\_k & Left-tail jump variation ($k_t$) & Bollerslev, Todorov \& Xu (2015) \\
65 & BTX\_MLJI\_10\_annual & Mean left-jump intensity (10\%) & Bollerslev, Todorov \& Xu (2015) \\
66 & BTX\_MLJI\_20\_annual & Mean left-jump intensity (20\%) & Bollerslev, Todorov \& Xu (2015) \\
67 & BTX\_MLJI\_7\_5\_annual & Mean left-jump intensity (7.5\%) & Bollerslev, Todorov \& Xu (2015) \\
68 & BTX\_MLJI\_k\_annual & Mean left-jump intensity ($k_t$) & Bollerslev, Todorov \& Xu (2015) \\
69 & BTX\_MRJI\_10\_annual & Mean right-jump intensity (10\%) & Bollerslev, Todorov \& Xu (2015) \\
70 & BTX\_MRJI\_20\_annual & Mean right-jump intensity (20\%) & Bollerslev, Todorov \& Xu (2015) \\
71 & BTX\_MRJI\_7\_5\_annual & Mean right-jump intensity (7.5\%) & Bollerslev, Todorov \& Xu (2015) \\
72 & BTX\_MRJI\_k\_annual & Mean right-jump intensity ($k_t$) & Bollerslev, Todorov \& Xu (2015) \\
73 & BTX\_phi\_minus & Left-tail intensity parameter $\phi_t^-$ & Bollerslev, Todorov \& Xu (2015) \\
74 & BTX\_phi\_plus & Right-tail intensity parameter $\phi_t^+$ & Bollerslev, Todorov \& Xu (2015) \\
75 & BTX\_PLJ\_10 & Probability of left jump (10\%) & Bollerslev, Todorov \& Xu (2015) \\
76 & BTX\_PLJ\_20 & Probability of left jump (20\%) & Bollerslev, Todorov \& Xu (2015) \\
77 & BTX\_PLJ\_7\_5 & Probability of left jump (7.5\%) & Bollerslev, Todorov \& Xu (2015) \\
78 & BTX\_PLJ\_k & Probability of left jump ($k_t$) & Bollerslev, Todorov \& Xu (2015) \\
79 & BTX\_PRJ\_10 & Probability of right jump (10\%) & Bollerslev, Todorov \& Xu (2015) \\
80 & BTX\_PRJ\_20 & Probability of right jump (20\%) & Bollerslev, Todorov \& Xu (2015) \\
81 & BTX\_PRJ\_7\_5 & Probability of right jump (7.5\%) & Bollerslev, Todorov \& Xu (2015) \\
82 & BTX\_PRJ\_k & Probability of right jump ($k_t$) & Bollerslev, Todorov \& Xu (2015) \\
83 & BTX\_RJV\_10 & Right-tail jump variation (10\%) & Bollerslev, Todorov \& Xu (2015) \\
84 & BTX\_RJV\_20 & Right-tail jump variation (20\%) & Bollerslev, Todorov \& Xu (2015) \\
85 & BTX\_RJV\_7\_5 & Right-tail jump variation (7.5\%) & Bollerslev, Todorov \& Xu (2015) \\
86 & BTX\_RJV\_k & Right-tail jump variation ($k_t$) & Bollerslev, Todorov \& Xu (2015) \\
87 & disaster\_RIX\_annual & RIX (annualized) & Gao, Gao \& Song (2018) \\
88 & disaster\_RIX\_call\_annual & RIX call (annualized) & Gao, Gao \& Song (2018) \\
89 & disaster\_RIX\_put\_annual & RIX put (annualized) & Gao, Gao \& Song (2018) \\
90 & implBeta & Option-implied beta &  \\
91 & implBeta\_chang & Option-implied beta from implied moments & Chang et al.~(2012) \\
92 & implERP\_beta\_annual & Option-implied ERP (beta-based) &  \\
93 & implERP\_SVIX2\_annual & Option-implied ERP (SVIX2-based) & Martin \& Wagner (2019) \\
94 & ivs\_atm\_spread & ATM IV call--put spread & Yan (2011) \\
95 & ivs\_convexity & IV convexity & Kelly et al.~(2016) \\
96 & ivs\_smirk & IV smirk & Yan (2011); Xing et al.~(2010) \\
97 & martinImplVol\_annual & SVIX implied volatility (annualized) & Martin (2017) \\
98 & oscore & Rank-average of IV, skewness, kurtosis & Alexiou \& Rompolis (2022) \\
99 & P\_implQ\_beta & Option-implied physical beta &  \\
100 & P\_implQ\_implVol\_annual & Option-implied physical volatility &  \\
101 & P\_implQ\_implVol\_eps\_annual & Physical idiosyncratic volatility &  \\
102 & P\_implQ\_implVol\_sys\_annual & Physical systematic volatility &  \\
103 & P\_implQ\_kurt & Physical kurtosis &  \\
104 & P\_implQ\_kurt\_eps & Physical idiosyncratic kurtosis &  \\
105 & P\_implQ\_kurt\_sys & Physical systematic kurtosis &  \\
106 & P\_implQ\_mu & Physical mean &  \\
107 & P\_implQ\_mu\_eps & Physical idiosyncratic mean &  \\
108 & P\_implQ\_mu\_sys & Physical systematic mean &  \\
109 & P\_implQ\_skew & Physical skewness &  \\
110 & P\_implQ\_skew\_eps & Physical idiosyncratic skewness &  \\
111 & P\_implQ\_skew\_sys & Physical systematic skewness &  \\
112 & P\_implQ\_var & Physical variance &  \\
113 & P\_implQ\_var\_eps & Physical idiosyncratic variance &  \\
114 & P\_implQ\_var\_sys & Physical systematic variance &  \\
115 & RND\_beta & RND beta &  \\
116 & RND\_implVol\_annual & RND implied volatility &  \\
117 & RND\_implVol\_eps\_annual & RND idiosyncratic volatility &  \\
118 & RND\_implVol\_sys\_annual & RND systematic volatility &  \\
119 & RND\_kurt & RND kurtosis &  \\
120 & RND\_kurt\_eps & RND idiosyncratic kurtosis &  \\
121 & RND\_kurt\_sys & RND systematic kurtosis &  \\
122 & RND\_mu & RND mean &  \\
123 & RND\_mu\_eps & RND idiosyncratic mean &  \\
124 & RND\_mu\_sys & RND systematic mean &  \\
125 & RND\_skew & RND skewness &  \\
126 & RND\_skew\_eps & RND idiosyncratic skewness &  \\
127 & RND\_skew\_sys & RND systematic skewness &  \\
128 & RND\_var & RND variance &  \\
129 & RND\_var\_eps & RND idiosyncratic variance &  \\
130 & RND\_var\_sys & RND systematic variance &  \\
131 & RVaR\_pct\_10 & Rescaled VaR Q (10\%) & Ammann \& Feser (2019) \\
132 & RVaR\_pct\_50 & Rescaled VaR Q (50\%) & Ammann \& Feser (2019) \\
133 & RVaR\_pct\_90 & Rescaled VaR Q (90\%) & Ammann \& Feser (2019) \\
134 & slope\_down & Put-wing slope & Kelly et al.~(2016) \\
135 & slope\_up & Call-wing slope (sign inverted) & Park et al.~(2019) \\
136 & SVIX2\_annual & Risk-neutral variance (annualized) & Martin (2017) \\
137 & SVIX2\_call\_annual & SVIX$^2$ call component & Martin (2017) \\
138 & SVIX2\_put\_annual & SVIX$^2$ put component & Martin (2017) \\
\end{longtable}
}
\clearpage

\section*{Appendix C: Robustness}
\label{sec:appendix_c}
\setcounter{table}{0}
\renewcommand{\thetable}{C.\arabic{table}}

This appendix reports the results summarized in Sections~\ref{sec:screen_results} and \ref{sec:robustness}: the stability of the screen, the DS-LASSO estimates at all maturities under both risk-free rates, and the aggregate tests at all maturities.

\subsection*{C.1 Screen Selection across Maturities}

Table~\ref{tab:gbr_persistence} reports, for each option maturity, whether the full-sample screen selects a factor. It lists every factor selected at least at one maturity.

\begin{small}
\begin{center}
\begin{longtable}{lcccccccc}
\caption{\textbf{Screen Selection across Maturities}} \label{tab:gbr_persistence} \\
\toprule
\textbf{Factor Identifier} & $\mathbf{\tau_1}$ & $\mathbf{\tau_2}$ & $\mathbf{\tau_3}$ & $\mathbf{\tau_5}$ & $\mathbf{\tau_{10}}$ & $\mathbf{\tau_{30}}$ & $\mathbf{\tau_{60}}$ & \textbf{Sum} \\
\midrule
\endfirsthead

\multicolumn{9}{c}{{\bfseries \tablename\ \thetable{} -- continued from previous page}} \\
\toprule
\textbf{Factor Identifier} & $\mathbf{\tau_1}$ & $\mathbf{\tau_2}$ & $\mathbf{\tau_3}$ & $\mathbf{\tau_5}$ & $\mathbf{\tau_{10}}$ & $\mathbf{\tau_{30}}$ & $\mathbf{\tau_{60}}$ & \textbf{Sum} \\
\midrule
\endhead

\bottomrule
\endfoot

\bottomrule
\multicolumn{9}{p{\textwidth}}{\vspace{2pt}\footnotesize This table reports whether the full-sample GBR screen selects each factor at option maturities $\tau$ of 1 to 60 days. Sum is the number of maturities at which the factor is selected.} \\
\endlastfoot

BTX\_MRJI\_k\_annual       & True  & True  & True  & True  & True  & True  & True  & 7 \\
BTX\_RJV\_k                & True  & True  & True  & True  & True  & True  & True  & 7 \\
BTX\_EQV\_relRightTail\_k  & True  & True  & True  & True  & True  & True  & False & 6 \\
BTX\_MLJI\_20\_annual      & True  & True  & True  & True  & True  & True  & False & 6 \\
ivs\_convexity             & True  & True  & True  & True  & True  & True  & False & 6 \\
BTX\_EQV\_relRightTail\_10 & True  & True  & True  & True  & False & True  & False & 5 \\
BTX\_EQV\_relTail\_10      & True  & True  & True  & True  & True  & False & False & 5 \\
BTX\_LJV\_7\_5             & True  & False & True  & True  & True  & True  & False & 5 \\
RND\_mu\_sys               & True  & True  & True  & True  & True  & False & False & 5 \\
bakshi\_X                  & True  & True  & False & True  & False & True  & True  & 5 \\
BTX\_LJV\_10               & False & True  & True  & False & True  & False & True  & 4 \\
bakshi\_X\_call            & True  & True  & True  & False & True  & False & False & 4 \\
implBeta\_chang            & True  & True  & False & False & True  & False & True  & 4 \\
BTX\_EQV\_relTail\_k       & True  & False & False & False & True  & True  & False & 3 \\
P\_implQ\_var              & True  & False & False & False & False & True  & True  & 3 \\
RND\_mu                    & True  & False & False & False & False & True  & True  & 3 \\
bakshiImplVar\_put         & False & False & True  & True  & True  & False & False & 3 \\
BJN\_V\_put\_annual        & False & False & False & True  & True  & False & False & 2 \\
BTX\_EQCV\_k               & False & False & False & False & False & True  & True  & 2 \\
BTX\_EQV\_relRightTail\_20 & False & False & False & True  & True  & False & False & 2 \\
BTX\_EQV\_relRightTail\_7\_5 & True & False & False & False & False & False & True  & 2 \\
BTX\_EQV\_relTail\_20      & True  & False & True  & False & False & False & False & 2 \\
BTX\_alpha\_plus           & False & False & False & False & False & True  & True  & 2 \\
P\_implQ\_var\_eps          & False & False & True  & False & False & True  & False & 2 \\
RND\_var\_sys              & False & True  & False & True  & False & False & False & 2 \\
SVIX2\_put\_annual         & True  & False & False & False & False & True  & False & 2 \\
BTX\_EQCV\_7\_5            & False & False & False & False & True  & False & False & 1 \\
BTX\_EQV\_relLeftTail\_20  & False & False & False & True  & False & False & False & 1 \\
BTX\_MLJI\_k\_annual       & True  & False & False & False & False & False & False & 1 \\
BTX\_MRJI\_20\_annual      & False & False & False & False & False & True  & False & 1 \\
BTX\_MRJI\_7\_5\_annual    & False & False & False & False & False & False & True  & 1 \\
BTX\_RJV\_10               & False & False & False & False & False & True  & False & 1 \\
P\_implQ\_kurt\_sys        & False & False & False & True  & False & False & False & 1 \\
RND\_kurt\_eps             & False & False & False & False & False & True  & False & 1 \\
RND\_skew\_sys             & False & False & False & False & True  & False & False & 1 \\
RND\_var                   & False & False & True  & False & False & False & False & 1 \\
bakshiImplVar             & False & False & False & False & True  & False & False & 1 \\
bakshiKurt                 & False & False & False & False & False & True  & False & 1 \\
bakshi\_W\_call            & False & False & False & False & False & False & True  & 1 \\
bakshi\_W\_put             & False & False & True  & False & False & False & False & 1 \\
bakshi\_X\_put             & False & True  & False & False & False & False & False & 1 \\
ivs\_smirk                 & False & False & False & False & False & True  & False & 1 \\
slope\_down                & False & False & False & False & False & False & True  & 1 \\
\end{longtable}
\end{center}
\end{small}

\clearpage

\subsection*{C.2 Out-of-Sample Fit of the Screen}

Table~\ref{tab:gbr_rmse} reports the root mean squared error (RMSE) of the GBR in the validation part of five expanding windows (EW1 to EW5) and five rolling windows (RW1 to RW5), for each option maturity. The target is the contemporaneous equal-weighted average test-asset return, so the RMSE measures out-of-sample fit rather than forecasting ability.

\begin{table}[H]
\centering
\begin{threeparttable}
\caption{\textbf{Out-of-Sample RMSE of the Screen}}
\label{tab:gbr_rmse}
\begin{small}
\begin{tabular}{lcccccccccc}
\toprule
$\mathbf{\tau}$ & \textbf{EW1} & \textbf{EW2} & \textbf{EW3} & \textbf{EW4} & \textbf{EW5} & \textbf{RW1} & \textbf{RW2} & \textbf{RW3} & \textbf{RW4} & \textbf{RW5} \\
\midrule
1  & 3.646 & 5.554 & 4.940 & 4.883 & 5.403 & 3.646 & 8.189 & 3.568 & 4.049 & 6.500 \\
2  & 3.649 & 5.525 & 4.931 & 4.904 & 5.430 & 3.649 & 8.192 & 3.565 & 4.049 & 6.683 \\
3  & 3.649 & 5.507 & 4.991 & 4.922 & 5.375 & 3.649 & 8.194 & 3.552 & 4.048 & 6.635 \\
5  & 3.645 & 5.565 & 4.899 & 4.902 & 5.405 & 3.645 & 8.195 & 3.550 & 4.049 & 6.585 \\
10 & 3.640 & 5.498 & 4.916 & 4.853 & 5.345 & 3.640 & 8.194 & 3.541 & 4.051 & 6.691 \\
30 & 3.631 & 5.513 & 4.902 & 4.856 & 5.347 & 3.631 & 8.191 & 3.541 & 4.053 & 6.641 \\
60 & 3.637 & 5.462 & 4.812 & 4.781 & 5.302 & 3.637 & 8.175 & 3.520 & 4.057 & 6.672 \\
\bottomrule
\end{tabular}
\end{small}
\begin{tablenotes}
\footnotesize \item RMSE in percent per month in the validation part of five expanding (EW) and five rolling (RW) windows, by option maturity $\tau$ in days.
\end{tablenotes}
\end{threeparttable}
\end{table}

\clearpage

\subsection*{C.3 Screen Selection in Subsamples}

Table~\ref{tab:gbr_window_selections} reports, for each option maturity, in how many of the five expanding windows (EW) and five rolling windows (RW) the screen selects a factor. Factors that are never selected are omitted.

\begin{small}
\begin{center}
\setlength{\tabcolsep}{3.5pt}
\begin{longtable}{lccccccccccccccc}
\caption{\textbf{Screen Selection in Subsamples}} \label{tab:gbr_window_selections} \\
\toprule
 & \multicolumn{7}{c}{\textbf{Expanding Windows (EW)}} & \multicolumn{7}{c}{\textbf{Rolling Windows (RW)}} & \\
\cmidrule(lr){2-8} \cmidrule(lr){9-15}
\textbf{Factor Identifier} & $\mathbf{\tau_1}$ & $\mathbf{\tau_2}$ & $\mathbf{\tau_3}$ & $\mathbf{\tau_5}$ & $\mathbf{\tau_{10}}$ & $\mathbf{\tau_{30}}$ & $\mathbf{\tau_{60}}$ & $\mathbf{\tau_1}$ & $\mathbf{\tau_2}$ & $\mathbf{\tau_3}$ & $\mathbf{\tau_5}$ & $\mathbf{\tau_{10}}$ & $\mathbf{\tau_{30}}$ & $\mathbf{\tau_{60}}$ & \textbf{Sum} \\
\midrule
\endfirsthead

\multicolumn{16}{c}{{\bfseries \tablename\ \thetable{} -- continued from previous page}} \\
\toprule
 & \multicolumn{7}{c}{\textbf{Expanding Windows (EW)}} & \multicolumn{7}{c}{\textbf{Rolling Windows (RW)}} & \\
\cmidrule(lr){2-8} \cmidrule(lr){9-15}
\textbf{Factor Identifier} & $\mathbf{\tau_1}$ & $\mathbf{\tau_2}$ & $\mathbf{\tau_3}$ & $\mathbf{\tau_5}$ & $\mathbf{\tau_{10}}$ & $\mathbf{\tau_{30}}$ & $\mathbf{\tau_{60}}$ & $\mathbf{\tau_1}$ & $\mathbf{\tau_2}$ & $\mathbf{\tau_3}$ & $\mathbf{\tau_5}$ & $\mathbf{\tau_{10}}$ & $\mathbf{\tau_{30}}$ & $\mathbf{\tau_{60}}$ & \textbf{Sum} \\
\midrule
\endhead

\bottomrule
\endfoot

\bottomrule
\multicolumn{16}{p{\textwidth}}{\vspace{2pt}\footnotesize Number of the five expanding (EW) and five rolling (RW) windows in which the screen selects the factor, by option maturity $\tau$ in days. Sum is the total over all 70 window-maturity combinations. Factors that are never selected are omitted.} \\
\endlastfoot

P\_implQ\_var               & 4 & 4 & 2 & 3 & 3 & 4 & 4 & 1 & 3 & 0 & 1 & 0 & 2 & 1 & 32 \\
BTX\_RJV\_k                & 3 & 3 & 3 & 3 & 3 & 3 & 3 & 1 & 1 & 1 & 1 & 1 & 1 & 1 & 28 \\
BTX\_RJV\_20               & 3 & 3 & 3 & 3 & 2 & 3 & 3 & 0 & 0 & 0 & 0 & 0 & 0 & 0 & 20 \\
implBeta\_chang            & 2 & 2 & 1 & 2 & 2 & 2 & 3 & 1 & 1 & 1 & 1 & 1 & 0 & 1 & 20 \\
BTX\_EQV\_relTail\_10      & 4 & 2 & 2 & 2 & 2 & 0 & 0 & 1 & 2 & 2 & 1 & 0 & 0 & 0 & 18 \\
P\_implQ\_var\_eps          & 1 & 2 & 4 & 2 & 2 & 2 & 0 & 1 & 1 & 1 & 1 & 1 & 0 & 0 & 18 \\
BTX\_MRJI\_k\_annual       & 2 & 2 & 2 & 2 & 2 & 2 & 2 & 1 & 0 & 1 & 1 & 0 & 0 & 1 & 18 \\
RND\_beta                  & 2 & 3 & 1 & 2 & 3 & 2 & 1 & 0 & 0 & 0 & 1 & 1 & 0 & 1 & 17 \\
implBeta                   & 0 & 1 & 2 & 3 & 3 & 3 & 0 & 1 & 0 & 0 & 1 & 1 & 1 & 1 & 17 \\
RND\_var\_sys              & 2 & 4 & 2 & 3 & 2 & 2 & 1 & 0 & 0 & 0 & 1 & 0 & 0 & 0 & 17 \\
BTX\_EQV\_relTail\_20      & 4 & 3 & 3 & 1 & 2 & 0 & 0 & 1 & 1 & 1 & 0 & 0 & 0 & 0 & 16 \\
BTX\_EQV\_relRightTail\_10 & 1 & 3 & 3 & 2 & 0 & 2 & 0 & 1 & 1 & 1 & 1 & 0 & 0 & 0 & 15 \\
BTX\_MLJI\_20\_annual      & 2 & 2 & 3 & 2 & 3 & 2 & 1 & 0 & 0 & 0 & 0 & 0 & 0 & 0 & 15 \\
BTX\_EQJV\_10              & 1 & 1 & 1 & 1 & 1 & 1 & 1 & 1 & 1 & 1 & 1 & 1 & 1 & 1 & 14 \\
BTX\_MRJI\_7\_5\_annual    & 1 & 1 & 1 & 1 & 2 & 1 & 2 & 0 & 0 & 1 & 2 & 1 & 1 & 0 & 14 \\
BTX\_LJV\_10               & 1 & 2 & 3 & 2 & 2 & 1 & 1 & 0 & 0 & 1 & 0 & 1 & 0 & 0 & 14 \\
BTX\_EQJV\_20              & 1 & 2 & 2 & 3 & 2 & 2 & 2 & 0 & 0 & 0 & 0 & 0 & 0 & 0 & 14 \\
BTX\_LJV\_7\_5             & 2 & 1 & 2 & 3 & 2 & 2 & 1 & 0 & 0 & 0 & 0 & 0 & 0 & 0 & 13 \\
BTX\_MLJI\_10\_annual      & 1 & 1 & 1 & 1 & 0 & 1 & 1 & 1 & 1 & 1 & 1 & 1 & 1 & 1 & 13 \\
BTX\_EQV\_relRightTail\_k  & 3 & 1 & 3 & 1 & 2 & 1 & 2 & 0 & 0 & 0 & 0 & 0 & 0 & 0 & 13 \\
bakshi\_X\_put             & 1 & 1 & 1 & 2 & 0 & 2 & 1 & 1 & 1 & 1 & 1 & 0 & 0 & 0 & 12 \\
BTX\_EQJV\_7\_5            & 1 & 1 & 1 & 2 & 2 & 2 & 2 & 0 & 0 & 0 & 0 & 0 & 0 & 0 & 11 \\
bakshi\_X                  & 1 & 1 & 0 & 3 & 1 & 1 & 1 & 0 & 0 & 0 & 0 & 0 & 1 & 2 & 11 \\
bakshiImplVar\_put         & 0 & 0 & 2 & 1 & 3 & 0 & 2 & 1 & 1 & 0 & 0 & 0 & 0 & 0 & 10 \\
RND\_mu\_sys               & 2 & 1 & 1 & 2 & 1 & 0 & 0 & 0 & 1 & 1 & 0 & 0 & 0 & 0 & 9 \\
BTX\_EQV\_relTail\_k       & 1 & 0 & 0 & 0 & 1 & 2 & 0 & 0 & 1 & 0 & 1 & 0 & 2 & 0 & 8 \\
BTX\_EQV\_relRightTail\_7\_5 & 1 & 0 & 0 & 0 & 0 & 1 & 1 & 1 & 1 & 1 & 1 & 1 & 0 & 0 & 8 \\
P\_implQ\_var\_sys          & 0 & 0 & 1 & 3 & 0 & 0 & 3 & 0 & 0 & 0 & 0 & 0 & 0 & 1 & 8 \\
RND\_mu                    & 1 & 0 & 0 & 0 & 0 & 1 & 3 & 0 & 1 & 0 & 0 & 0 & 0 & 2 & 8 \\
ivs\_convexity             & 1 & 1 & 1 & 1 & 2 & 1 & 0 & 0 & 0 & 0 & 0 & 0 & 0 & 0 & 7 \\
bakshi\_X\_call            & 1 & 1 & 1 & 0 & 1 & 0 & 1 & 0 & 0 & 0 & 0 & 1 & 0 & 1 & 7 \\
BTX\_EQV\_relTail\_7\_5      & 1 & 0 & 0 & 1 & 0 & 1 & 0 & 1 & 0 & 0 & 1 & 1 & 0 & 1 & 7 \\
ivs\_smirk                 & 1 & 1 & 0 & 0 & 1 & 2 & 0 & 0 & 0 & 0 & 0 & 0 & 1 & 0 & 6 \\
disaster\_RIX\_annual       & 0 & 0 & 0 & 0 & 0 & 1 & 4 & 0 & 0 & 0 & 0 & 0 & 0 & 1 & 6 \\
bakshi\_W                  & 0 & 0 & 0 & 0 & 0 & 3 & 0 & 0 & 0 & 0 & 1 & 0 & 2 & 0 & 6 \\
SVIX2\_put\_annual         & 2 & 0 & 0 & 1 & 0 & 1 & 0 & 1 & 0 & 0 & 0 & 0 & 1 & 0 & 6 \\
BTX\_EQV\_relRightTail\_20 & 0 & 0 & 0 & 3 & 2 & 0 & 0 & 0 & 0 & 0 & 0 & 0 & 0 & 0 & 5 \\
BTX\_EQCV\_7\_5            & 0 & 0 & 0 & 0 & 1 & 3 & 0 & 0 & 0 & 0 & 0 & 1 & 0 & 0 & 5 \\
disaster\_RIX\_put\_annual   & 0 & 0 & 0 & 0 & 0 & 0 & 3 & 0 & 0 & 1 & 0 & 0 & 0 & 1 & 5 \\
BTX\_EQCV\_10              & 0 & 1 & 2 & 0 & 0 & 0 & 2 & 0 & 0 & 0 & 0 & 0 & 0 & 0 & 5 \\
BTX\_EQCV\_k               & 1 & 0 & 0 & 0 & 0 & 2 & 1 & 0 & 0 & 0 & 0 & 0 & 0 & 1 & 5 \\
bakshi\_W\_put             & 0 & 0 & 1 & 1 & 0 & 1 & 0 & 0 & 0 & 0 & 0 & 0 & 1 & 0 & 4 \\
P\_implQ\_mu\_eps          & 0 & 0 & 1 & 0 & 0 & 0 & 0 & 1 & 1 & 1 & 0 & 0 & 0 & 0 & 4 \\
RND\_mu\_eps               & 1 & 0 & 1 & 0 & 1 & 1 & 0 & 0 & 0 & 0 & 0 & 0 & 0 & 0 & 4 \\
slope\_down                & 0 & 0 & 0 & 1 & 0 & 0 & 1 & 0 & 0 & 0 & 0 & 1 & 0 & 1 & 4 \\
disaster\_RIX\_call\_annual  & 0 & 0 & 0 & 1 & 2 & 0 & 0 & 0 & 0 & 0 & 0 & 0 & 0 & 0 & 3 \\
BTX\_EQCV\_20              & 0 & 0 & 0 & 0 & 0 & 0 & 3 & 0 & 0 & 0 & 0 & 0 & 0 & 0 & 3 \\
RND\_var\_eps              & 1 & 0 & 0 & 0 & 0 & 0 & 1 & 0 & 0 & 1 & 0 & 0 & 0 & 0 & 3 \\
BTX\_EQV\_relLeftTail\_k   & 0 & 0 & 0 & 0 & 0 & 0 & 0 & 0 & 1 & 2 & 0 & 0 & 0 & 0 & 3 \\
BTX\_LJV\_20               & 0 & 0 & 0 & 0 & 0 & 0 & 0 & 0 & 0 & 1 & 0 & 1 & 1 & 0 & 3 \\
bakshiKurt                 & 0 & 0 & 0 & 0 & 0 & 3 & 0 & 0 & 0 & 0 & 0 & 0 & 0 & 0 & 3 \\
RVaR\_pct\_50              & 0 & 0 & 0 & 0 & 2 & 0 & 0 & 0 & 0 & 0 & 0 & 1 & 0 & 0 & 3 \\
RND\_var                   & 0 & 1 & 2 & 0 & 0 & 0 & 0 & 0 & 0 & 0 & 0 & 0 & 0 & 0 & 3 \\
RND\_kurt\_eps             & 0 & 0 & 0 & 0 & 1 & 1 & 1 & 0 & 0 & 0 & 0 & 0 & 0 & 0 & 3 \\
RND\_skew\_eps             & 0 & 0 & 0 & 0 & 0 & 1 & 0 & 0 & 0 & 0 & 0 & 0 & 1 & 0 & 2 \\
bakshiImplVar              & 0 & 0 & 1 & 0 & 1 & 0 & 0 & 0 & 0 & 0 & 0 & 0 & 0 & 0 & 2 \\
BJN\_V\_put\_annual        & 0 & 0 & 0 & 1 & 1 & 0 & 0 & 0 & 0 & 0 & 0 & 0 & 0 & 0 & 2 \\
P\_implQ\_kurt\_sys        & 0 & 0 & 1 & 1 & 0 & 0 & 0 & 0 & 0 & 0 & 0 & 0 & 0 & 0 & 2 \\
BTX\_alpha\_plus           & 0 & 0 & 0 & 0 & 0 & 1 & 1 & 0 & 0 & 0 & 0 & 0 & 0 & 0 & 2 \\
RND\_skew\_sys             & 0 & 0 & 0 & 0 & 1 & 0 & 0 & 0 & 0 & 0 & 0 & 0 & 0 & 0 & 1 \\
BTX\_MRJI\_20\_annual      & 0 & 0 & 0 & 0 & 0 & 1 & 0 & 0 & 0 & 0 & 0 & 0 & 0 & 0 & 1 \\
BTX\_MLJI\_k\_annual       & 1 & 0 & 0 & 0 & 0 & 0 & 0 & 0 & 0 & 0 & 0 & 0 & 0 & 0 & 1 \\
oscore                     & 0 & 0 & 0 & 1 & 0 & 0 & 0 & 0 & 0 & 0 & 0 & 0 & 0 & 0 & 1 \\
ivs\_atm\_spread           & 0 & 0 & 0 & 0 & 0 & 0 & 0 & 0 & 0 & 0 & 0 & 0 & 0 & 1 & 1 \\
BJN\_V\_call\_annual       & 0 & 0 & 0 & 0 & 1 & 0 & 0 & 0 & 0 & 0 & 0 & 0 & 0 & 0 & 1 \\
BTX\_EQJV\_k               & 0 & 0 & 0 & 0 & 0 & 0 & 0 & 0 & 0 & 0 & 1 & 0 & 0 & 0 & 1 \\
BTX\_EQV\_relLeftTail\_10  & 0 & 0 & 0 & 0 & 0 & 0 & 0 & 0 & 0 & 1 & 0 & 0 & 0 & 0 & 1 \\
BTX\_EQV\_relLeftTail\_20  & 0 & 0 & 0 & 1 & 0 & 0 & 0 & 0 & 0 & 0 & 0 & 0 & 0 & 0 & 1 \\
bakshiSkew                 & 0 & 0 & 0 & 0 & 0 & 0 & 0 & 0 & 0 & 0 & 0 & 1 & 0 & 0 & 1 \\
bakshi\_W\_call            & 0 & 0 & 0 & 0 & 0 & 0 & 1 & 0 & 0 & 0 & 0 & 0 & 0 & 0 & 1 \\
RND\_skew                  & 0 & 0 & 0 & 0 & 0 & 0 & 0 & 0 & 0 & 0 & 0 & 0 & 0 & 1 & 1 \\
BTX\_RJV\_10               & 0 & 0 & 0 & 0 & 0 & 1 & 0 & 0 & 0 & 0 & 0 & 0 & 0 & 0 & 1 \\
BTX\_alpha\_minus          & 0 & 0 & 0 & 0 & 0 & 0 & 0 & 0 & 0 & 0 & 0 & 1 & 0 & 0 & 1 \\
bakshiImplVar\_call        & 0 & 0 & 0 & 0 & 0 & 0 & 0 & 0 & 0 & 0 & 0 & 0 & 0 & 1 & 1 \\
bakshiImplVar\_asym        & 0 & 0 & 1 & 0 & 0 & 0 & 0 & 0 & 0 & 0 & 0 & 0 & 0 & 0 & 1 \\
SVIX2\_call\_annual        & 0 & 0 & 0 & 0 & 0 & 0 & 0 & 0 & 0 & 0 & 1 & 0 & 0 & 0 & 1 \\
SVIX2\_annual              & 0 & 0 & 0 & 0 & 1 & 0 & 0 & 0 & 0 & 0 & 0 & 0 & 0 & 0 & 1 \\
slope\_up                  & 0 & 0 & 0 & 0 & 0 & 0 & 1 & 0 & 0 & 0 & 0 & 0 & 0 & 0 & 1 \\
\end{longtable}
\end{center}
\end{small}

\clearpage

\subsection*{C.4 DS-LASSO Estimates across Maturities: FRED Risk-Free Rate}

Table~\ref{tab:ds_lasso_fred} reports the DS-LASSO estimates for all factors selected at each maturity ($\tau \in \{1, 2, 3, 5, 10, 30, 60\}$ days), using the FRED Treasury-bill rate as in the main text. The 30-day column corresponds to Table~\ref{tab:ds_results}.

\begin{small}
\begin{center}
\setlength{\tabcolsep}{3.5pt}
\newcolumntype{M}{>{\centering\arraybackslash}p{1.3cm}}

\begin{longtable}{lMMMMMMM}
\caption{\textbf{DS-LASSO Estimates across Maturities (FRED)}} \label{tab:ds_lasso_fred} \\
\toprule
\textbf{Factor Identifier} & $\mathbf{\tau_1}$ & $\mathbf{\tau_2}$ & $\mathbf{\tau_3}$ & $\mathbf{\tau_5}$ & $\mathbf{\tau_{10}}$ & $\mathbf{\tau_{30}}$ & $\mathbf{\tau_{60}}$ \\
\midrule
\endfirsthead

\multicolumn{8}{c}{{\bfseries \tablename\ \thetable{} -- continued from previous page}} \\
\toprule
\textbf{Factor Identifier} & $\mathbf{\tau_1}$ & $\mathbf{\tau_2}$ & $\mathbf{\tau_3}$ & $\mathbf{\tau_5}$ & $\mathbf{\tau_{10}}$ & $\mathbf{\tau_{30}}$ & $\mathbf{\tau_{60}}$ \\
\midrule
\endhead

\bottomrule
\endfoot

\bottomrule
\multicolumn{8}{p{\textwidth}}{\vspace{2pt}\footnotesize Each cell reports $\lambda_s$ in basis points per month and, in parentheses, its $t$-statistic from the DS-LASSO test of the factor against the 160 equity factors at option maturity $\tau$ in days. -- indicates that the factor is not selected by the screen at that maturity. Excess returns use the FRED Treasury-bill rate.} \\
\endlastfoot

BJN\_V\_put\_annual         & -- & -- & -- & 104 (1.02) & -30 (-0.46) & -- & -- \\
BTX\_EQCV\_7\_5            & -- & -- & -- & -- & -290 (-3.01) & -- & -- \\
BTX\_EQCV\_k               & -- & -- & -- & -- & -- & -34 (-0.30) & -182 (-2.25) \\
BTX\_EQV\_relLeftTail\_20  & -- & -- & -- & 492 (5.94) & -- & -- & -- \\
BTX\_EQV\_relRightTail\_10 & 43 (0.40) & 62 (2.03) & 393 (3.54) & -57 (-0.60) & -- & -55 (-0.39) & -- \\
BTX\_EQV\_relRightTail\_20 & -- & -- & -- & 32 (0.50) & -24 (-0.14) & -- & -- \\
BTX\_EQV\_relRightTail\_7\_5 & -104 (-0.34) & -- & -- & -- & -- & -- & -120 (-2.30) \\
BTX\_EQV\_relRightTail\_k  & 150 (2.99) & 155 (2.23) & 49 (1.60) & 201 (2.72) & 55 (2.31) & -128 (-1.82) & -- \\
BTX\_EQV\_relTail\_10      & -234 (-2.67) & -175 (-2.13) & -314 (-3.57) & -1723 (-0.77) & 2090 (2.36) & -- & -- \\
BTX\_EQV\_relTail\_20      & 23 (0.32) & -- & 393 (4.70) & -- & -- & -- & -- \\
BTX\_EQV\_relTail\_k       & 265 (1.99) & -- & -- & -- & 175 (3.42) & 295 (3.65) & -- \\
BTX\_LJV\_10               & -- & 274 (4.84) & 90 (1.66) & -- & 117 (2.31) & -- & -296 (-2.38) \\
BTX\_LJV\_7\_5             & 122 (1.97) & -- & 73 (3.09) & 282 (2.62) & 73 (2.16) & 446 (3.59) & -- \\
BTX\_MLJI\_20\_annual      & -195 (-1.71) & 72 (0.62) & -85 (-0.70) & -382 (-4.36) & -15 (-0.21) & -171 (-1.93) & -- \\
BTX\_MLJI\_k\_annual       & 104 (0.80) & -- & -- & -- & -- & -- & -- \\
BTX\_MRJI\_20\_annual      & -- & -- & -- & -- & -- & -310 (-2.28) & -- \\
BTX\_MRJI\_7\_5\_annual    & -- & -- & -- & -- & -- & -- & 419 (2.87) \\
BTX\_MRJI\_k\_annual       & 88 (2.81) & 292 (3.25) & 128 (2.61) & 155 (2.87) & 47 (0.25) & 66 (0.79) & 102 (1.51) \\
BTX\_RJV\_10               & -- & -- & -- & -- & -- & 44 (1.76) & -- \\
BTX\_RJV\_k                & 26 (0.16) & -89 (-1.26) & 24 (0.34) & 0 (0.01) & -201 (-2.27) & 116 (1.79) & 380 (2.57) \\
BTX\_alpha\_plus           & -- & -- & -- & -- & -- & 74 (2.23) & -31 (-0.45) \\
P\_implQ\_kurt\_sys        & -- & -- & -- & 193 (1.77) & -- & -- & -- \\
P\_implQ\_var              & 91 (3.46) & -- & -- & -- & -- & -261 (-2.40) & 43 (1.93) \\
P\_implQ\_var\_eps          & -- & -- & -97 (-0.53) & -- & -- & 63 (0.69) & -- \\
RND\_kurt\_eps             & -- & -- & -- & -- & -- & 151 (2.89) & -- \\
RND\_mu                    & 560 (7.01) & -- & -- & -- & -- & -141 (-1.32) & 114 (2.40) \\
RND\_mu\_sys               & -91 (-0.83) & 52 (2.22) & 461 (3.77) & -225 (-2.61) & -20 (-0.22) & -- & -- \\
RND\_skew\_sys             & -- & -- & -- & -- & 98 (0.98) & -- & -- \\
RND\_var                   & -- & -- & -238 (-2.06) & -- & -- & -- & -- \\
RND\_var\_sys              & -- & 323 (1.95) & -- & 77 (3.19) & -- & -- & -- \\
SVIX2\_put\_annual         & -259 (-2.73) & -- & -- & -- & -- & 60 (0.40) & -- \\
bakshiImplVar              & -- & -- & -- & -- & 379 (2.65) & -- & -- \\
bakshiImplVar\_put         & -- & -- & -208 (-2.17) & -93 (-0.90) & 9 (0.10) & -- & -- \\
bakshiKurt                 & -- & -- & -- & -- & -- & 79 (3.15) & -- \\
bakshi\_W\_call            & -- & -- & -- & -- & -- & -- & -336 (-3.61) \\
bakshi\_W\_put             & -- & -- & 22 (0.32) & -- & -- & -- & -- \\
bakshi\_X                  & 67 (0.50) & -258 (-2.41) & -- & 142 (1.95) & -- & 65 (0.98) & 345 (4.13) \\
bakshi\_X\_call            & -70 (-0.87) & -240 (-2.42) & -115 (-1.10) & -- & -198 (-1.90) & -- & -- \\
bakshi\_X\_put             & -- & -53 (-0.52) & -- & -- & -- & -- & -- \\
implBeta\_chang            & 139 (1.52) & -64 (-0.48) & -- & -- & 320 (3.85) & -- & -132 (-1.32) \\
ivs\_convexity             & 130 (2.00) & 124 (2.57) & 113 (1.10) & 73 (1.53) & -163 (-1.95) & 420 (3.92) & -- \\
ivs\_smirk                 & -- & -- & -- & -- & -- & -45 (-0.88) & -- \\
slope\_down                & -- & -- & -- & -- & -- & -- & 66 (0.74) \\
\end{longtable}
\end{center}
\end{small}

\clearpage

\subsection*{C.5 DS-LASSO Estimates across Maturities: TOPT Risk-Free Rate}
Table~\ref{tab:ds_lasso_topt} repeats Table~\ref{tab:ds_lasso_fred} with excess returns based on the option-implied TOPT rate \citep{vanbinsbergen2022risk, vanbinsbergen2023options}. The risk-free rate cancels out of the long--short factors, and a common shift in the test assets' excess returns is absorbed by the intercept of the cross-sectional regression. Differences between the two tables therefore arise mainly because the LASSO steps select slightly different controls. Where the selected controls coincide, the estimates are identical.

\begin{small}
\begin{center}
\setlength{\tabcolsep}{3.5pt}
\newcolumntype{M}{>{\centering\arraybackslash}p{1.3cm}}

\begin{longtable}{lMMMMMMM}
\caption{\textbf{DS-LASSO Estimates across Maturities (TOPT)}} \label{tab:ds_lasso_topt} \\
\toprule
\textbf{Factor Identifier} & $\mathbf{\tau_1}$ & $\mathbf{\tau_2}$ & $\mathbf{\tau_3}$ & $\mathbf{\tau_5}$ & $\mathbf{\tau_{10}}$ & $\mathbf{\tau_{30}}$ & $\mathbf{\tau_{60}}$ \\
\midrule
\endfirsthead

\multicolumn{8}{c}{{\bfseries \tablename\ \thetable{} -- continued from previous page}} \\
\toprule
\textbf{Factor Identifier} & $\mathbf{\tau_1}$ & $\mathbf{\tau_2}$ & $\mathbf{\tau_3}$ & $\mathbf{\tau_5}$ & $\mathbf{\tau_{10}}$ & $\mathbf{\tau_{30}}$ & $\mathbf{\tau_{60}}$ \\
\midrule
\endhead

\bottomrule
\endfoot

\bottomrule
\multicolumn{8}{p{\textwidth}}{\vspace{2pt}\footnotesize Each cell reports $\lambda_s$ in basis points per month and, in parentheses, its $t$-statistic from the DS-LASSO test of the factor against the 160 equity factors at option maturity $\tau$ in days. -- indicates that the factor is not selected by the screen at that maturity. Excess returns use the option-implied TOPT rate.} \\
\endlastfoot

BJN\_V\_put\_annual         & -- & -- & -- & 104 (1.02) & 42 (0.62) & -- & -- \\
BTX\_EQCV\_7\_5            & -- & -- & -- & -- & -290 (-3.01) & -- & -- \\
BTX\_EQCV\_k               & -- & -- & -- & -- & -- & -5 (-0.04) & -146 (-1.60) \\
BTX\_EQV\_relLeftTail\_20  & -- & -- & -- & 498 (5.94) & -- & -- & -- \\
BTX\_EQV\_relRightTail\_10 & 32 (0.29) & 105 (3.26) & 393 (3.54) & -57 (-0.60) & -- & 182 (1.29) & -- \\
BTX\_EQV\_relRightTail\_20 & -- & -- & -- & 25 (0.39) & -24 (-0.14) & -- & -- \\
BTX\_EQV\_relRightTail\_7\_5 & -147 (-0.47) & -- & -- & -- & -- & -- & -111 (-2.22) \\
BTX\_EQV\_relRightTail\_k  & 150 (2.99) & 170 (2.43) & 42 (1.39) & 185 (2.48) & 67 (2.83) & -128 (-1.84) & -- \\
BTX\_EQV\_relTail\_10      & -234 (-2.67) & -236 (-2.65) & -407 (-4.67) & -2873 (-1.19) & 1708 (2.03) & -- & -- \\
BTX\_EQV\_relTail\_20      & -15 (-0.21) & -- & 388 (4.69) & -- & -- & -- & -- \\
BTX\_EQV\_relTail\_k       & 120 (0.86) & -- & -- & -- & 178 (3.43) & 295 (3.65) & -- \\
BTX\_LJV\_10               & -- & 244 (4.39) & 90 (1.66) & -- & 105 (2.11) & -- & -296 (-2.38) \\
BTX\_LJV\_7\_5             & 144 (2.77) & -- & 77 (3.25) & 282 (2.62) & 68 (2.09) & 446 (3.59) & -- \\
BTX\_MLJI\_20\_annual      & -166 (-1.42) & 72 (0.62) & -104 (-0.80) & -382 (-4.36) & -10 (-0.13) & -149 (-1.76) & -- \\
BTX\_MLJI\_k\_annual       & 104 (0.80) & -- & -- & -- & -- & -- & -- \\
BTX\_MRJI\_20\_annual      & -- & -- & -- & -- & -- & -310 (-2.28) & -- \\
BTX\_MRJI\_7\_5\_annual    & -- & -- & -- & -- & -- & -- & 419 (2.87) \\
BTX\_MRJI\_k\_annual       & 57 (1.95) & 293 (3.38) & 98 (2.01) & 116 (2.04) & 47 (0.25) & 72 (0.85) & 102 (1.51) \\
BTX\_RJV\_10               & -- & -- & -- & -- & -- & 44 (1.76) & -- \\
BTX\_RJV\_k                & -160 (-1.14) & -190 (-2.52) & 24 (0.34) & 0 (0.01) & -201 (-2.27) & 116 (1.87) & 363 (2.48) \\
BTX\_alpha\_plus           & -- & -- & -- & -- & -- & 74 (2.23) & -30 (-0.46) \\
P\_implQ\_kurt\_sys        & -- & -- & -- & 193 (1.77) & -- & -- & -- \\
P\_implQ\_var              & 71 (2.97) & -- & -- & -- & -- & -340 (-2.77) & 84 (3.24) \\
P\_implQ\_var\_eps          & -- & -- & -159 (-0.85) & -- & -- & 63 (0.69) & -- \\
RND\_kurt\_eps             & -- & -- & -- & -- & -- & 151 (2.89) & -- \\
RND\_mu                    & 546 (6.81) & -- & -- & -- & -- & -169 (-1.58) & 114 (2.40) \\
RND\_mu\_sys               & -91 (-0.83) & 52 (2.22) & 461 (3.77) & -220 (-2.39) & -20 (-0.22) & -- & -- \\
RND\_skew\_sys             & -- & -- & -- & -- & 94 (0.95) & -- & -- \\
RND\_var                   & -- & -- & -238 (-2.06) & -- & -- & -- & -- \\
RND\_var\_sys              & -- & 369 (2.24) & -- & 77 (3.19) & -- & -- & -- \\
SVIX2\_put\_annual         & -275 (-2.87) & -- & -- & -- & -- & 60 (0.40) & -- \\
bakshiImplVar              & -- & -- & -- & -- & 93 (0.59) & -- & -- \\
bakshiImplVar\_put         & -- & -- & -208 (-2.17) & -93 (-0.90) & 9 (0.10) & -- & -- \\
bakshiKurt                 & -- & -- & -- & -- & -- & 79 (3.15) & -- \\
bakshi\_W\_call            & -- & -- & -- & -- & -- & -- & -336 (-3.61) \\
bakshi\_W\_put             & -- & -- & 22 (0.32) & -- & -- & -- & -- \\
bakshi\_X                  & 67 (0.50) & -293 (-2.64) & -- & 142 (1.95) & -- & 89 (1.31) & 330 (3.94) \\
bakshi\_X\_call            & -118 (-1.51) & -239 (-2.44) & -115 (-1.10) & -- & -198 (-1.90) & -- & -- \\
bakshi\_X\_put             & -- & -48 (-0.47) & -- & -- & -- & -- & -- \\
implBeta\_chang            & 139 (1.45) & 4 (0.04) & -- & -- & 311 (3.78) & -- & -132 (-1.32) \\
ivs\_convexity             & 100 (1.44) & 125 (2.59) & 113 (1.10) & 73 (1.53) & -115 (-1.35) & 420 (3.92) & -- \\
ivs\_smirk                 & -- & -- & -- & -- & -- & -45 (-0.88) & -- \\
slope\_down                & -- & -- & -- & -- & -- & -- & 66 (0.74) \\
\end{longtable}
\end{center}
\end{small}

\clearpage

\subsection*{C.6 GRS Tests across Maturities}

Table~\ref{tab:grs_robustness} reports GRS tests of whether the alphas of the selected option factors relative to the 160 equity factors are jointly zero, for each option maturity. The test does not reject at the 5\% level at six of the seven maturities. At 60 days, it rejects with $p = 0.043$. With seven maturities, one rejection at the 5\% level is not unusual even if the null hypothesis is true at all of them.

\begin{table}[H]
\centering
\begin{threeparttable}
\caption{\textbf{GRS Tests across Maturities}}
\label{tab:grs_robustness}
\begin{small}
\setlength{\tabcolsep}{7pt}
\begin{tabular}{lccccccc}
\toprule
 & $\mathbf{\tau_1}$ & $\mathbf{\tau_2}$ & $\mathbf{\tau_3}$ & $\mathbf{\tau_5}$ & $\mathbf{\tau_{10}}$ & $\mathbf{\tau_{30}}$ & $\mathbf{\tau_{60}}$ \\
\midrule
GRS statistic & 1.105 & 1.197 & 1.557 & 1.477 & 1.631 & 1.246 & 1.918 \\
                   & (0.3683) & (0.2979) & (0.1067) & (0.1356) & (0.0787) & (0.2500) & (0.0427) \\
\bottomrule
\end{tabular}
\end{small}
\begin{tablenotes}
\footnotesize \item GRS statistics for the null hypothesis that the alphas of the option factors selected at each maturity $\tau$ (in days) relative to the 160 equity factors are jointly zero. $p$-values are in parentheses.
\end{tablenotes}
\end{threeparttable}
\end{table}

\bigskip

\subsection*{C.7 Maximum Sharpe Ratios across Maturities}

Table~\ref{tab:spanning_robustness} reports in-sample maximum Sharpe ratios of the equity factors, the option factors, and both combined, for each option maturity. The option factors alone reach between 0.85 and 1.29. Adding them to the equity factors raises the in-sample maximum Sharpe ratio by 0.16 to 0.39. These values are biased upward, and the increase is not statistically significant except at 60 days (Table~\ref{tab:grs_robustness}).

\begin{table}[H]
\centering
\begin{threeparttable}
\caption{\textbf{Maximum Sharpe Ratios across Maturities}}
\label{tab:spanning_robustness}
\begin{small}
\setlength{\tabcolsep}{5.5pt}
\begin{tabular}{lccccccc}
\toprule
 & $\mathbf{\tau_1}$ & $\mathbf{\tau_2}$ & $\mathbf{\tau_3}$ & $\mathbf{\tau_5}$ & $\mathbf{\tau_{10}}$ & $\mathbf{\tau_{30}}$ & $\mathbf{\tau_{60}}$ \\
\midrule
Equity factors & 3.347 & 3.347 & 3.347 & 3.347 & 3.347 & 3.347 & 3.347 \\
Option factors & 1.040 & 0.853 & 1.040 & 1.100 & 1.039 & 1.293 & 0.887 \\
Combined & 3.577 & 3.508 & 3.632 & 3.585 & 3.592 & 3.738 & 3.558 \\
Increase (combined $-$ equity) & 0.2300 & 0.1608 & 0.2843 & 0.2374 & 0.2442 & 0.3906 & 0.2107 \\
\bottomrule
\end{tabular}
\end{small}
\begin{tablenotes}
\footnotesize \item Annualized in-sample maximum Sharpe ratios of the tangency portfolios of the 160 equity factors, the option factors selected at each maturity $\tau$ (in days), and both combined, February 2004 to July 2023. Covariance matrices use the shrinkage estimator of \citet{ledoit2004well}.
\end{tablenotes}
\end{threeparttable}
\end{table}

\bigskip

\subsection*{C.8 Redundancy across Maturities}

Table~\ref{tab:cca_robustness} reports the redundancy indices of Table~\ref{tab:cca_results} for each option maturity. At every maturity, the equity factors explain between 91\% and 94\% of the variance of the option factors, and the option factors explain between 29\% and 35\% of the variance of the equity factors. The approximate noise levels are 0.69 and 0.09.

\begin{table}[H]
\centering
\begin{threeparttable}
\caption{\textbf{Redundancy across Maturities}}
\label{tab:cca_robustness}
\begin{small}
\setlength{\tabcolsep}{4.5pt}
\begin{tabular}{lccccccc}
\toprule
 & $\mathbf{\tau_1}$ & $\mathbf{\tau_2}$ & $\mathbf{\tau_3}$ & $\mathbf{\tau_5}$ & $\mathbf{\tau_{10}}$ & $\mathbf{\tau_{30}}$ & $\mathbf{\tau_{60}}$ \\
\midrule
Options explained by equity factors ($RI_{O|E}$) & 0.9271 & 0.9303 & 0.9291 & 0.9218 & 0.9221 & 0.9103 & 0.9359 \\
Equity factors explained by options ($RI_{E|O}$) & 0.3097 & 0.2876 & 0.2885 & 0.3068 & 0.3128 & 0.3536 & 0.3291 \\
\midrule
Unique share of options ($1 - RI_{O|E}$) & 0.0729 & 0.0697 & 0.0709 & 0.0782 & 0.0779 & 0.0897 & 0.0641 \\
Unique share of equity factors ($1 - RI_{E|O}$) & 0.6903 & 0.7124 & 0.7115 & 0.6932 & 0.6872 & 0.6464 & 0.6709 \\
\bottomrule
\end{tabular}
\end{small}
\begin{tablenotes}
\footnotesize \item Stewart--Love redundancy indices between the option factors selected at each maturity $\tau$ (in days) and the 160 equity factors, estimated as in Table~\ref{tab:cca_results}.
\end{tablenotes}
\end{threeparttable}
\end{table}

\end{appendices}
\end{document}